\documentclass[aps,prd,twocolumn,superscriptaddress,preprintnumbers]{revtex4}

\usepackage[latin1]{inputenc}
\usepackage{epsfig,graphicx,color}
\usepackage{amsmath}
\usepackage{amsfonts}
\usepackage{amssymb}
\usepackage{rotating}
\usepackage{subfigure}

\newcommand{\be}{\begin{equation}}
\newcommand{\ee}{\end{equation}}
\newcommand{\bea}{\begin{eqnarray}}
\newcommand{\eea}{\end{eqnarray}}
\newcommand{\f}[2]{\ensuremath{\frac{#1}{#2}}}
\newcommand{\bef}{\begin{figure}[htbp]\begin{center}}
\newcommand{\eef}{\end{center}\end{figure}}

\newcommand{\gsim}{\lower.7ex\hbox{$\;\stackrel{\textstyle>}{\sim}\;$}}
\newcommand{\lsim}{\lower.7ex\hbox{$\;\stackrel{\textstyle<}{\sim}\;$}}

\newcommand{\Lann}{\mathcal{L}_{\rm ann}}
\begin{document}

\preprint{SLAC-PUB-14200}

\title{Indirect Dark Matter Detection Limits from the 
\\ Ultra-Faint Milky Way Satellite Segue 1}

\author{Rouven Essig}
\affiliation{SLAC National Accelerator Laboratory, Stanford, California, 94309, USA}

\author{Neelima Sehgal}
\affiliation{Kavli Institute for Particle Astrophysics and Cosmology, Stanford University, Stanford, California 94305, USA}

\author{Louis E. Strigari}
\affiliation{Kavli Institute for Particle Astrophysics and Cosmology, Stanford University, Stanford, California 94305, USA}

\author{Marla Geha}
\affiliation{Astronomy Department, Yale University, New Haven, CT 06520, USA}

\author{Joshua D. Simon}
\affiliation{Observatories of the Carnegie Institution of Washington, 813 Santa Barbara Street, Pasadena, CA 91101, USA}

\date{\today}

\begin{abstract}
We use new kinematic data from the ultra-faint Milky Way satellite Segue 1
to model its dark matter distribution and derive 
upper limits on the dark matter annihilation cross-section. 
Using gamma-ray flux upper limits from the Fermi satellite and MAGIC, we determine cross-section 
exclusion regions for dark matter annihilation into a variety of different particles including charged leptons.
We show that these exclusion regions are beginning to probe the regions of interest for a dark matter 
interpretation of the electron and positron fluxes from
PAMELA, Fermi, and HESS, and that future observations of Segue 1 have strong prospects for testing such an interpretation.  
We additionally discuss prospects for detecting annihilation with neutrinos using 
the IceCube detector, finding that in an optimistic scenario a few neutrino events may
be detected. Finally we use the kinematic data to model the Segue 1 dark matter velocity dispersion and 
constrain Sommerfeld enhanced models. 
\end{abstract}

\maketitle

\section{Introduction}\label{sec:intro} 

The Fermi Gamma-ray Space Telescope and the current 
generation of Imaging Atmospheric Cherenkov Telescopes (ACTs) 
such as VERITAS, MAGIC, HESS, and CANGAROO, 
are exploring the gamma-ray sky to unprecedented accuracy.  
Additionally, neutrino detectors such as IceCube will be probing the neutrino 
sky with greater sensitivity, lower energy thresholds, and finer angular resolution than achieved to date.
If weakly interacting massive particles (WIMPs) constitute the dominant component of the dark matter
in the Universe, experiments such as Fermi, ACTs, and IceCube may be 
sensitive to their annihilation or decay products.

Dark matter dominated Milky Way (MW) satellite galaxies are excellent targets to search for dark matter 
annihilation products, because they are nearby and largely free
of astrophysically-produced high energy photons. Among the population of MW satellites, of
particular interest is the ultra-low luminosity object Segue 1, which 
was discovered in 2006 as an
overdensity of resolved stars in the Sloan Digital Sky Survey (SDSS)~\cite{Belokurov:2006ph}
at a heliocentric distance of $23\pm2$ kpc. 
Originally classified as a tidally-disrupted globular cluster, follow-up spectroscopy for
individual stars has revealed that Segue 1 is the least luminous and most dark-matter dominated
galaxy known~\cite{Geha:2008zr}. Though its proximity to the Sagittarius stream has led to
a suggestion that Segue 1 is a dissolved star cluster originally associated with the 
Sagittarius dwarf galaxy~\cite{NiedersteOstholt:2009na}, new measurements of Segue 1 
member stars strongly support that it is indeed a dark matter dominated object 
\cite{Simon:2010ek,Geha2009}. 

Given its location at $23\pm2$ kpc and the fact that it is dark matter dominated, 
Segue 1 is an ideal source for dark matter studies~\cite{Geha:2008zr,Simon:2010ek}.
Since Segue 1 can be easily localized, any signal or limit on high energy gamma-ray/neutrino
emission from it is more straightforward to interpret than a corresponding diffuse signal from the Milky Way halo.
As we show in this paper, the integral over the square of the dark matter density distribution is similar to, 
and potentially larger than, the same quantity for all other known Milky Way satellites.

In this work we give the cross-section exclusion regions for dark matter annihilation into a variety of Standard Model particles using gamma-ray flux upper limits for Segue 1.  
We also give exclusion regions for dark matter annihilation to charged leptons, where an 
accompanying gamma-ray signal is guaranteed to exist from 
``final-state radiation'' (FSR) \cite{Bergstrom:2004cy,Beacom:2004pe,Birkedal2005}.  
The resulting flux from FSR photons is independent of 
the astrophysical environment in which the photons are produced (as opposed
to inverse Compton or synchrotron processes). 
Neutrino production is also guaranteed from the decay of muon or tau leptons, and 
we also investigate the case in which dark matter annihilates purely into neutrinos. 

Constraints on dark matter annihilation to charged leptons are
particularly relevant in light of the positron results from 
the PAMELA satellite in the energy range of 10-100 GeV~\cite{Adriani2008} 
as well as the Fermi \cite{Abdo:2009zk} 
and HESS \cite{Collaboration:2008aaa,Aharonian:2009ah} 
$e^+ + e^-$ spectra at $\sim 0.3 - 1$ TeV.  
If dark matter annihilation is responsible 
for these measurements, the annihilation cross-section must be 
$\mathcal{O}(100-1000)$ times larger than the thermal freeze-out cross-section 
and only a small number of anti-protons can be created in the annihilation 
process~\cite{Adriani:2008zq,Cirelli:2008pk,Arkani-Hamed2008, Pospelov:2008jd,Cholis:2008qq,Cholis2008,Bergstrom:2009fa,Meade:2009iu}. 
Given the importance of the dark matter interpretation of these measurements, and the fact that more precise 
$e^+$/$e^-$ flux measurements
may not conclusively determine their origin 
(given the uncertainty in the $e^+$/$e^-$ propagation in the Galaxy and 
in their background fluxes), independent checks of the dark matter interpretation of the PAMELA, Fermi, and
HESS signals are of the utmost importance. 

Here we present annihilation cross-section constraints from the 9-month
Fermi gamma-ray flux upper limits for Segue 1.  We also present constraints from gamma-ray flux 
upper limits obtained by MAGIC.  If dark matter annihilation
is responsible for the electron and/or positron signals observed by PAMELA and/or Fermi/HESS, 
we also show that the ACTs MAGIC and VERITAS 
have excellent prospects for testing models and detecting dark matter by observations of Segue 1. 
Finally, we investigate the prospects for detecting a neutrino signal from dark matter annihilation
in Segue 1 with IceCube, and we give constraints on Sommerfeld enhanced models.  
Regarding Sommerfeld models that explain both Fermi and PAMELA, we find that 
current limits slightly disfavor light mediators of mass $\mathcal{O}$(0.1 GeV) 
if we take the mean value for the Segue 1 line-of-sight integral of the dark matter halo 
density-squared.  

\section{Flux Calculation}\label{sec:flux}

The gamma-ray/neutrino flux from annihilating dark matter is 
\begin{equation}\label{equ:flux}
\frac{dN_{\imath}}{dAdt} = \frac{1}{8\pi} \mathcal{L}_{\rm{ann}}(\rho^{2}(r),D)
\frac{\langle\sigma v\rangle}{m_{\chi}^2} \int_{E_{th}}^{E_{max}}
\frac{dN_{\imath}}{dE_{\imath}} dE_{\imath}
\end{equation}
where $m_{\chi}$  is the dark matter particle mass, $\langle\sigma v\rangle$ is the annihilation cross-section, 
$E_{th}$ is the threshold energy of a given gamma-ray/neutrino instrument, and $E_{max}$ is the maximum energy of the 
photons/neutrinos.  The integral over $\frac{dN_{\imath}}{dE_{\imath}} $ depends only on the particle physics details of 
the dark matter annihilation.  $\mathcal{L}_{\rm{ann}}$ is the line-of-sight integral over the 
square of the dark matter density
\begin{equation}
 \mathcal{L}_{\rm{ann}} = \int_{0}^{\Delta\Omega}
\left\{  \int_{LOS} \rho^{2}(r) ds \right\} d\Omega
\label{eq:LOS}
\end{equation}
and depends only on the properties of the dark matter halo and the solid angle over which 
it is observed.
Here $\rho(r)$ is the halo dark matter density profile, $r = \sqrt{D^2+s^2-2sD\rm{cos}\theta}$ for a halo at a 
distance $D$, and the integral is performed along the line of sight over a solid angle $\Delta \Omega = 2\pi (1-
\cos \theta)$.  Here we assume $\langle\sigma v\rangle$ is constant throughout the halo.

{\bf Dark Matter Distribution} --- 
To determine $\mathcal{L}_{\rm{ann}}$ 
we use the new sample of line-of-sight velocities of Segue~1 member stars from~\cite{Simon:2010ek,Geha2009}. 
This sample consists of 71 probable member stars (as compared to the 
sample of 24 members used in previous indirect detection 
studies~\cite{Geha:2008zr,Martinez:2009jh,Essig:2009jx,Scott:2009jn,Perelstein:2010at}). 
In our calculations we include 66 stars, excluding from the sample of 71 
two RR Lyrae variable stars and two probable binaries identified in \cite{Simon:2010ek}, 
and also excluding the star SDSSJ100704.35+160459.4 which is a $6\sigma$ outlier from 
the mean velocity of Segue 1 \footnote{We note that including the 
two RR Lyrae variable stars and two probable binaries has minimal effect on our results.}.  
(See \cite{Simon:2010ek} for a detailed discussion of these stars and
the sample in general, including the effects of binary star contamination
~\cite{Minor:2010vp}.) We adopt the maximum likelihood procedure that is described
in detail in \cite{Strigari:2007at,Martinez:2009jh}, and review the 
aspects of the calculation that are important for
the purposes of this paper. We assume a gaussian likelihood
in the distribution of stellar velocities, with a dispersion given by the quadrature sum of the 
measurement uncertainty of each stellar velocity and the intrinsic dispersion, 
the latter of which is given by a solution to the Jeans equation. The Jeans equation
takes as input the dark matter density profile, stellar density profile, and the stellar velocity 
anisotropy. We assume that the dark matter and stellar density profiles are independent. 
For the dark matter density profile we assume a three-parameter Einasto profile, 
$\rho(r) = \rho_s \exp\Big\{ {-\frac{2}{\alpha}\Big[\Big(\frac{r}{r_s}\Big)^\alpha - 1 \Big]}\Big\}$.
A profile of this form is motivated by numerical simulations, which suggest that 
$\alpha=0.17$ provides a good fit for dark matter
halos over a wide variety of mass scales~\cite{Navarro:2008kc}. (Though we use the
Einasto parameterization in this paper, the results we present would not change if 
we use a different parameterization, e.g.~the generalized NFW model considered
in~\cite{Essig:2009jx}.) 
For simplicity in this paper we consider isotropic models with $\beta=0$;
we find that this restriction has minimal impact on our resulting constraints. 
The Segue 1 three-dimensional stellar density profile used in the Jeans equation 
is taken as a Plummer profile, $\rho(r) \propto (1+r^2/r_p^2)^{-5/2}$, where $r_p=30$ pc.  
We take uniform priors on the following parameters over the following ranges: 
$\alpha \in [0.14:0.3]$, $r_s/{\rm kpc} \in [0.01:0.6]$, and 
$\log_{10} \left[\rho_s/({\rm M}_\odot {\rm kpc}^{-3})\right] \in [6:10]$. We then marginalize 
over the gaussian likelihood to construct the likelihood for these parameters. 
Note that given our assumption of uniform priors the likelihood is equivalent to the posterior 
probability density distribution for the parameters.   

\begin{table}[t]
\begin{ruledtabular}
\begin{center}
\begin{tabular}{cc}
Within a radius of $\theta$ & $\mathcal{L}_{\rm ann}$ \\
(degree) & $\log_{10}[\mathcal{L}_{\rm{ann}}/(\rm{GeV}^2 \rm{cm}^{-5})]$ \\
\hline
\hline
0.01 & $18.3 \pm 0.9$ \\
0.1   & $19.0  \pm 0.6$\\
0.25   & $19.1 \pm 0.6$\\
1.0   & $19.2 \pm 0.6$\\ 
\end{tabular}
\caption{Segue 1 $\mathcal{L}_{\rm ann}$ values within different angular regions, 
with $\theta = 0.25^\circ$ corresponding to the radius of the outermost confirmed member star.
}
\label{tab:SegLs}
\end{center}
\end{ruledtabular}
\end{table}

From the likelihood for the parameters $\rho_s$, r$_s$, and $\alpha$, we construct the 
likelihood for ${\cal L}_{\rm ann}$ using Eq.~(\ref{eq:LOS}). To remain conservative we integrate the 
density squared within a solid angle of $\theta = 0.25^\circ$, corresponding to 100 pc.  
This is close to the radius of the outermost known bound star in the Segue 1 population, 
which lies at 87 pc. 
We fit gaussians to the log of the 
${\cal L}_{\rm ann}$ likelihood, and for Segue 1 we find 
\begin{equation}\label{eq:Segue1}
\mathcal{L}_{\rm{ann}}^{\rm Segue 1} = 10^{19.1 \pm 0.6} \, \mbox{GeV}^2\,{\rm cm}^{-5}, 
\end{equation}
with 1$\sigma$ error bars and $\theta$ within $0.25^\circ$.  
See Table~\ref{tab:SegLs} for ${\cal L}_{\rm ann}$ values of Segue 1 within other radii.

For the calculations below of the expected Sommerfeld enhancement, 
we are also interested in the velocity dispersion of the dark matter. Segue 1
is a particularly interesting, if not the most interesting, galaxy to consider 
in the context of Sommerfeld enhanced models because of its extremely low velocity dispersion. 
In our calculations we will consider the one-dimensional radial dark matter 
velocity dispersion, $\sigma_r$, averaged over the three-dimensional
dark matter mass distribution. 
We define this quantity as
\begin{equation}
\langle \sigma_r^2 \rangle = \frac{\int \rho(r) \sigma_r^2 r^2 dr}{\int \rho(r) r^2 dr}, 
\label{eq:sigma_dm}
\end{equation}  
where $\rho(r)$ is the Einasto profile as above. 
From a given density profile, the one-dimensional dispersion
$\sigma_r^2$ is derived from the Jeans equation, assuming that the dark 
matter velocity dispersion is isotropic. The likelihood for this quantity may then be
determined from our base set of three parameters in a manner similar to the derivation
of ${\cal L}_{\rm ann}$.  
We find
\be
\sqrt{\langle \sigma_r^2 \rangle} \simeq 4.5 \pm 1.8\;{\rm km/s}.
\ee
We stress again that this quantity is the radial velocity dispersion of the dark matter, which we find to
have a mean that is slightly larger than the observed line-of-sight velocity of stars reported in 
Ref.~\cite{Simon:2010ek}. 

The value of $\mathcal{L}_{\rm{ann}}^{\rm Segue 1}$ can be compared with a canonical value 
from the Galactic center.  
Using an Einasto density profile with standard Milky Way fit 
parameters, $\rho_s = 0.06$ GeV/cm$^3$ and $r_s = 20$ kpc ~\cite{Regis:2009qt}, we find 
$\mathcal{L}_{\rm{ann}}^{\rm MW} = 10^{20.82} \, \mbox{GeV}^2\,{\rm cm}^{-5}$ 
within a radius of $0.25^\circ$.
Although this is about two orders of magnitude larger than the mean value of 
$\mathcal{L}_{\rm{ann}}^{\rm Segue 1}$ within the same solid angle, the large astrophysical 
background in the Galactic center makes Segue 1 a much cleaner target.  
Moreover, the Milky Way density profile is estimated from simulations that 
have not included the effect of baryons, whose 
impact on the density profile is not clear.   
In addition, these simulations assume that the dark matter is cold (e.g.~a thermal WIMP or an axion), 
but a non-thermal WIMP, or a WIMP with non-negligible self-interactions, could give 
different halo profiles \cite{Colin2000}. 
Though the Milky Way dark matter distribution can be probed by stellar kinematics 
on larger scales (e.g. \cite{Widrow2005}), 
the distribution near the Galactic Center (within less than $\sim 100$s of pc), where 
most of the annihilation flux emanates, is currently completely unknown.
 
The mean value of $\Lann$ for Segue 1 is the same as for Draco \cite{Abdo:2010ex}, and is 
larger than that for any other known dwarf.  Although previous studies have derived larger 
$\Lann$ values for Willman 1 and Sagittarius \cite{Essig:2009jx}, neither system appears to be 
in dynamical equilibrium \cite{Willman:2010gy,Ibata:1995fz}, which makes constraints on their 
dark matter halos from existing data unreliable. 
 
{\bf Gamma-rays from Dark Matter} ---
For the direct annihilation channels into leptons,  $\chi\chi \to \ell\bar{\ell}$, where $\ell$ is a lepton, 
the energy spectrum from final state radiation (FSR) is given by
\begin{equation}
\frac{dN_{\gamma}}{dy}
=
\frac{\alpha}{\pi}\left(\frac{1 + (1 - y)^2}{y}\right)\left(\ln\bigg(\frac{s (1 - y)}{m_{\ell}^2}\bigg) - 1 \right)
\label{equ:fsr1}
\end{equation}
\citep{Beacom:2004pe,Birkedal2005,Mack:2008wu,Bell2008,Cholis2008,Fortin:2009rq}.
Here $\alpha\simeq 1/137$, $y=E_{\gamma}/m_{\chi}$, $s=4 m_{\chi}^2$, and $E_{max}=m_\chi$ in Eq.~(\ref{equ:flux}).
This formula holds in the collinear limit, where the photon is emitted collinearly with one
of the leptons and $m_\ell\ll m_\chi$.

For the channels that include $\tau$'s, the photon contribution from $\tau$ decays into $\pi^0$'s (which then  
decay to $\gamma$'s) dominates over the FSR signal, and the energy spectrum is given in 
\cite{Fornengo:2004kj,Essig:2009jx}.
The energy spectrum for FSR for the channel $\chi\chi \to \phi\phi \to \ell\bar{\ell}\ell\bar{\ell}$ is given 
explicitly in \cite{Essig:2009jx}.  
If the final state includes $\mu$'s, we also include photons from the radiative decay of the 
muon, e.g.~$\mu^{-} \to e^- \nu_\mu \bar{\nu}_e \gamma$, using the formulas found in e.g.~\cite{Kuno:1999jp,Mardon2009,Essig:2009jx}.
We do not consider gamma-ray signals from synchrotron radiation or inverse Compton scattering as both are expected to be negligible 
in Segue 1. 

\begin{figure*}[!t]
\begin{center}     
\includegraphics[width=.485\textwidth]{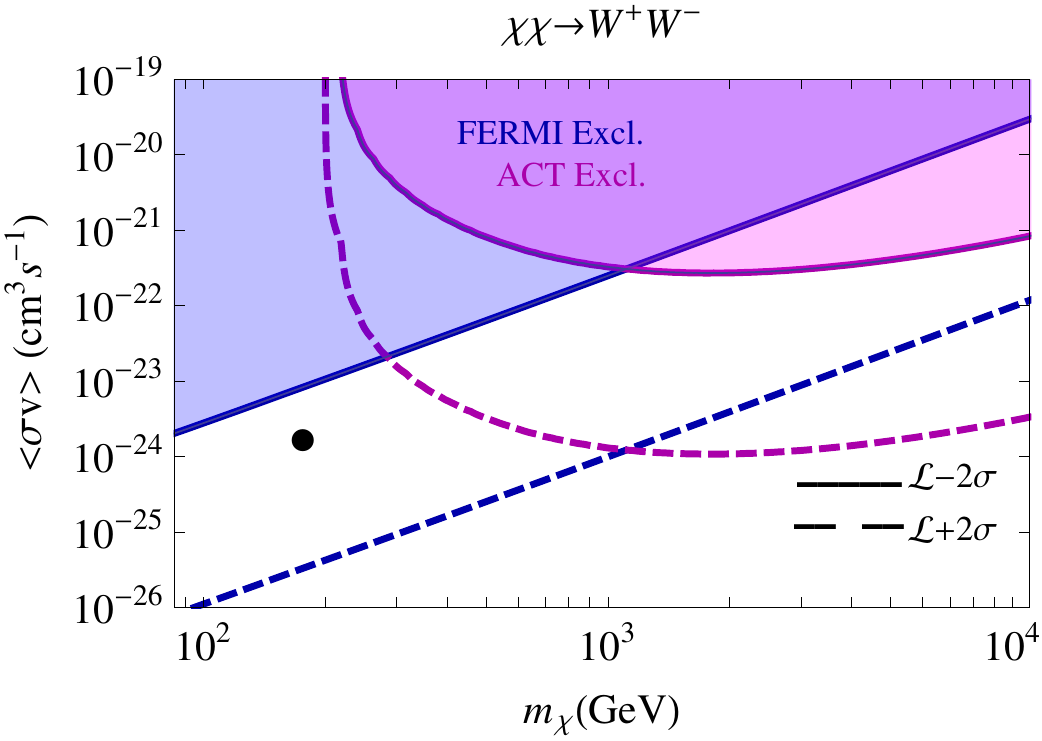}
\;
\includegraphics[width=.485\textwidth]{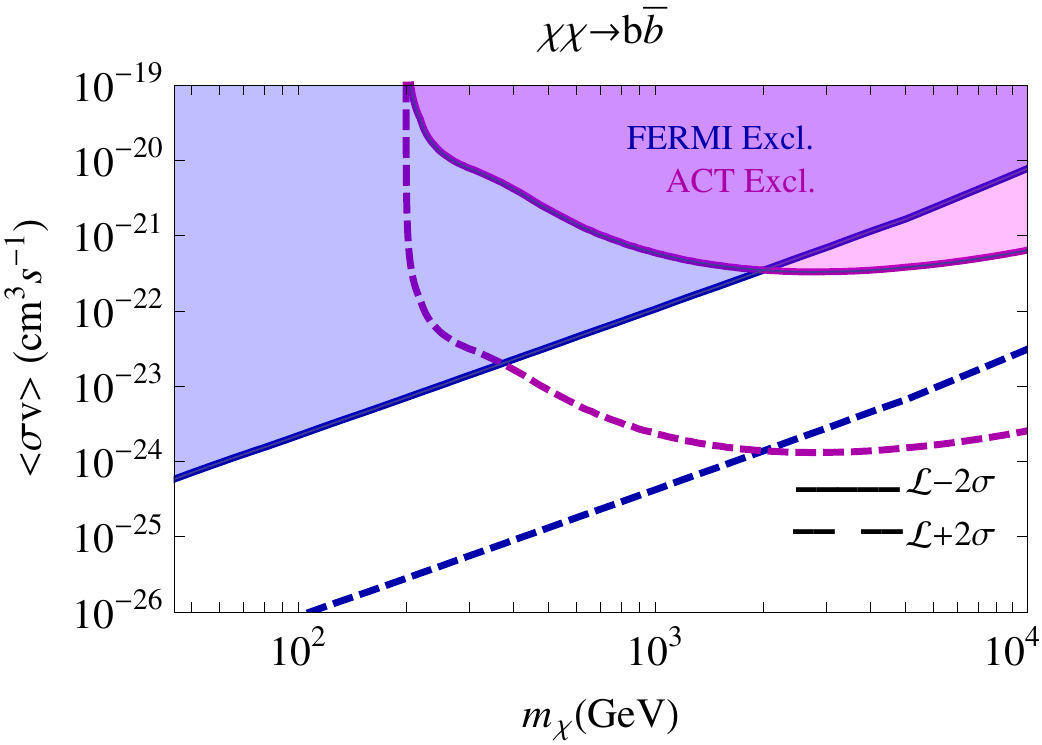}\\
\vskip 5mm
\includegraphics[width=.485\textwidth]{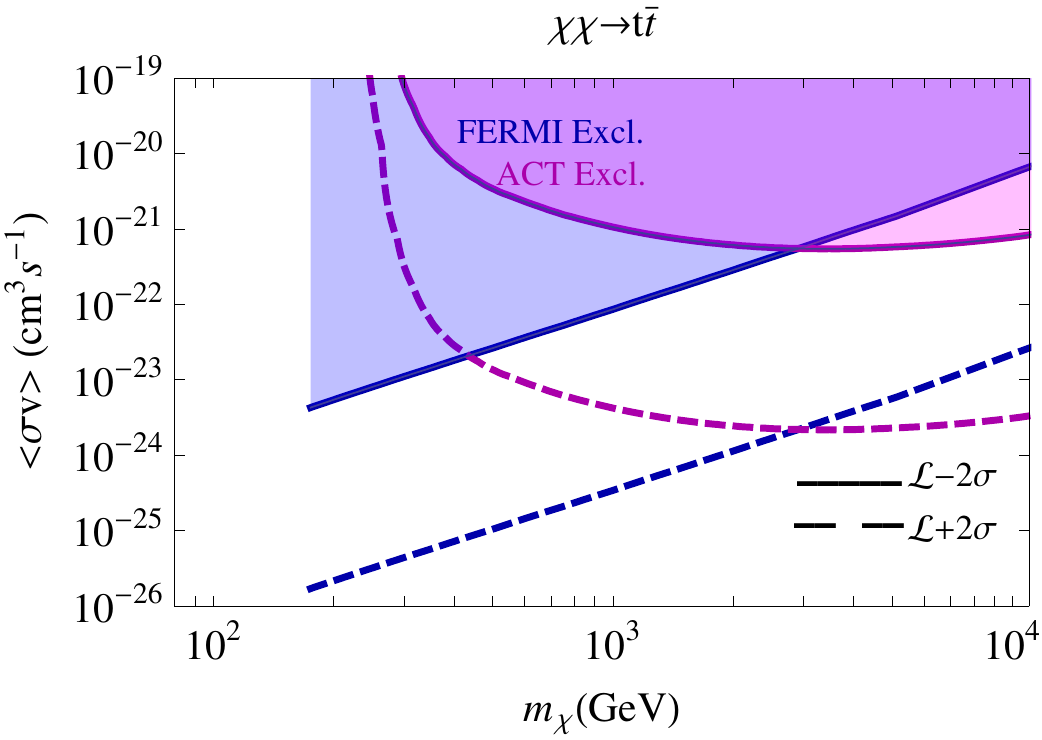}
\;
\includegraphics[width=.485\textwidth]{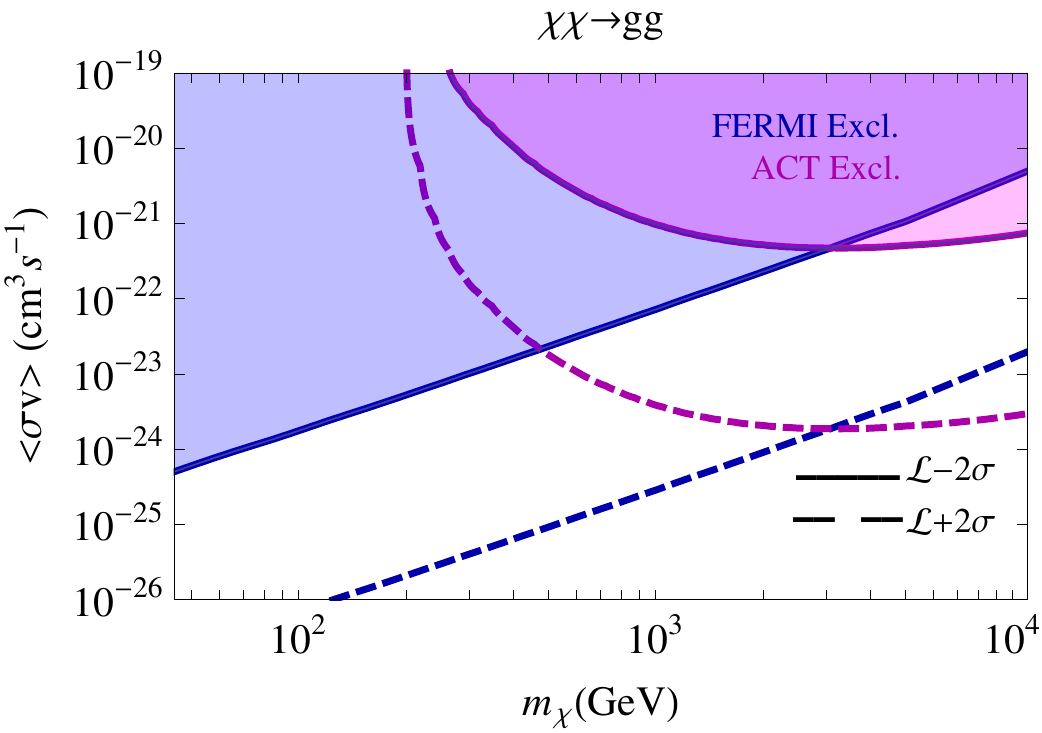}
\caption{Current exclusion regions from Fermi 9-month gamma-ray observations of Segue 1 (bounded below by the blue solid diagonal line) and MAGIC gamma-ray observations of Segue 1 (bounded below by the purple solid curved line).  
The exclusion regions use the conservative $2\sigma$ lower limit of $\mathcal{L}_{\rm{ann}}$ given in Table \ref{tab:SegLs} within $\theta = 0.25$ (Fermi) and $\theta=0.1$ (MAGIC).
The dashed blue and purple lines depict the respective 
cross-section bounds using the optimistic $2\sigma$ upper 
limits of $\mathcal{L}_{\rm{ann}}$.
For the $\chi\chi\to W^+W^-$ channel, the black dot is the region favored by a model of 
wino-like neutralinos that explains the PAMELA positron data \cite{Kane:2009if}.
Note that $m_\chi\gtrsim m_{\rm t} \simeq 175$ GeV for the $\chi\chi\to t\bar{t}$ channel.} 
\label{fig:gammabounds0}
\end{center}
\end{figure*}

For dark matter annihilation into $W^+W^-$ ($W$ bosons), $b\bar{b}$ (bottom quarks), 
$t\bar{t}$ (top quarks), and $gg$ (gluons), 
we use the photon spectra from DarkSUSY \cite{Gondolo:2004sc,darksusy}.  
Photons produced from these annihilation channels tend to be softer than FSR photons, but more 
numerous.  This makes Fermi more sensitive than ACTs to these annihilation channels.
For example, for the Wino LSP model \cite{Kane:2009if} that explains the PAMELA $e^+$ and 
$\bar{p}$ data (but not the Fermi or HESS $e^++e^-$ data), 
the dominant channel 
is $\chi\chi\to W^+W^-$, and the photon contribution comes from 
$W$ decays generating $\pi^0$'s and hence $\gamma$'s. 
To compare with Fermi flux upper limits, the relevant integral over  $dN_{\gamma}/dE_\gamma$ for this model is given by 
$\int_{100\,{\rm MeV}}^{m_\chi} (dN_\gamma/dE_\gamma) \, dE_\gamma$ $\simeq 27.14$,
where $m_\chi \simeq 180$ GeV \cite{Kane:2009if}.  
Since most of the photons have energies around 1 GeV,  ACTs will not 
be very constraining for this model even if they lower 
their energy threshold to $\sim 50$ GeV.  

{\bf Neutrinos from Dark Matter} --- We estimate the number of muons detected by IceCube
from muon neutrinos coming from
dark matter annihilation in Segue 1 following
e.g.~\cite{Barger:2007xf,Mardon2009} (see also \cite{Sandick:2009bi}).
We consider the direct channel $\chi\chi\to \nu_\mu \bar{\nu}_\mu$, which gives a 
good indication of how well IceCube could do in a ``best-case'' scenario, as well as the 
two channels $\chi\chi\to \mu^+\mu^-$ and
$\chi\chi\to \phi\phi$ with $\phi \to \mu^+\mu^-$ (we do not consider channels with taus, 
although their decay would also produce neutrinos).  
The muon decays $\mu^- \to \gamma\nu_\mu \bar\nu_e e^-$ and $\mu^+ \to
\gamma\bar\nu_\mu \nu_e e^+$ produce both
electron and muon (anti-)neutrinos.
The neutrino energy spectra for the direct channel
$\chi\chi\to \nu_\mu \bar{\nu}_\mu$ is given by 
\be
\frac{dN_{\nu}}{dE_\nu} = 2 \delta(E_\nu-m_\chi), 
\ee
while for the channel $\chi\chi\to \mu^+\mu^-$ we have 
\begin{eqnarray}
\frac{dN_{\nu_e}}{dx} & = & 2 - 6 x^2 + 4 x^3 \\
\frac{dN_{\nu_\mu}}{dx} & = & \frac{5}{3} - 3 x^2 + \frac{4}{3} x^3.
\end{eqnarray}
The energy spectra for the longer channel $\chi\chi\to \phi\phi$
with $\phi \to \mu^+\mu^-$
are given by
\begin{eqnarray}
\frac{dN_{\nu_e}}{dx} & = & -\frac{5}{3} + 3 x^2 - \frac{4}{3} x^3 + 2
\ln \frac{1}{x} \\
\frac{dN_{\nu_\mu}}{dx} & = & -\frac{19}{18} + \frac{3}{2} x^2 -
\frac{4}{9} x^3 + \frac{5}{3} \ln\frac{1}{x},
\end{eqnarray}
where $x=E_\nu/m_\chi$ \cite{Mardon2009}.  (The $\nu_i$ and $\bar\nu_i$ spectra are
the same for each channel).
When these neutrinos reach Earth, the probability that $\nu_i$ will have oscillated into $\nu_\mu$ is 
roughly \cite{Amsler:2008zzb} 
\begin{eqnarray}\label{eq:neutrinoosc}
P(\nu_\mu\to \nu_\mu) & \simeq & P(\bar\nu_\mu\to \bar\nu_\mu) \simeq 0.39, \nonumber \\ 
P(\nu_e\to \nu_\mu) & \simeq & P(\bar\nu_e\to \bar\nu_\mu) \simeq 0.22
\end{eqnarray}
(we take this oscillation probability also into account for the direct channel 
$\chi\chi\to \nu_\mu \bar{\nu}_\mu$). 
Since Segue 1 lies in the Northern Hemisphere, these neutrinos travel
through the Earth towards IceCube.
While $\nu_e$ and $\nu_\tau$ predominantly give rise to cascade-like events in
IceCube, $\nu_\mu$'s can convert to muons in the ice and produce
track-like events that yield much better angular resolution \cite{Resconi:2008fe}.  
We thus focus exclusively on detecting the muons from the $\nu_\mu$'s.  
(In particular, we also ignore the muons produced from $\nu_\tau \to \tau \to \mu$.)

\begin{figure*}[!t]
\begin{center}     
\includegraphics[width=.485\textwidth]{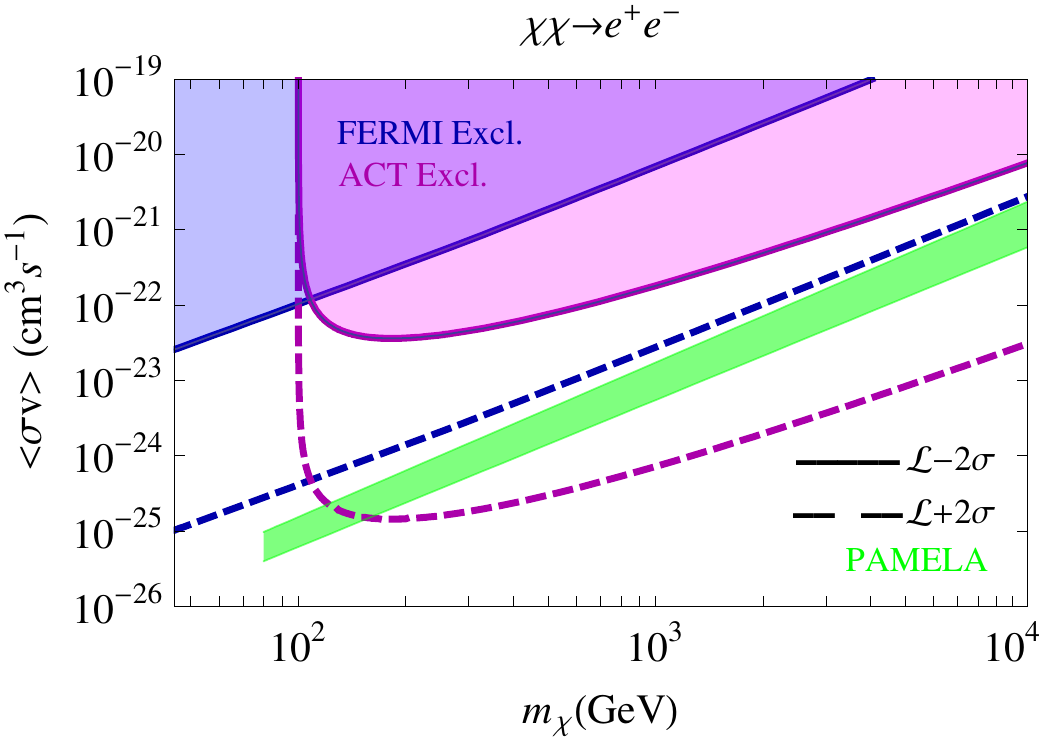}
\;
\includegraphics[width=.485\textwidth]{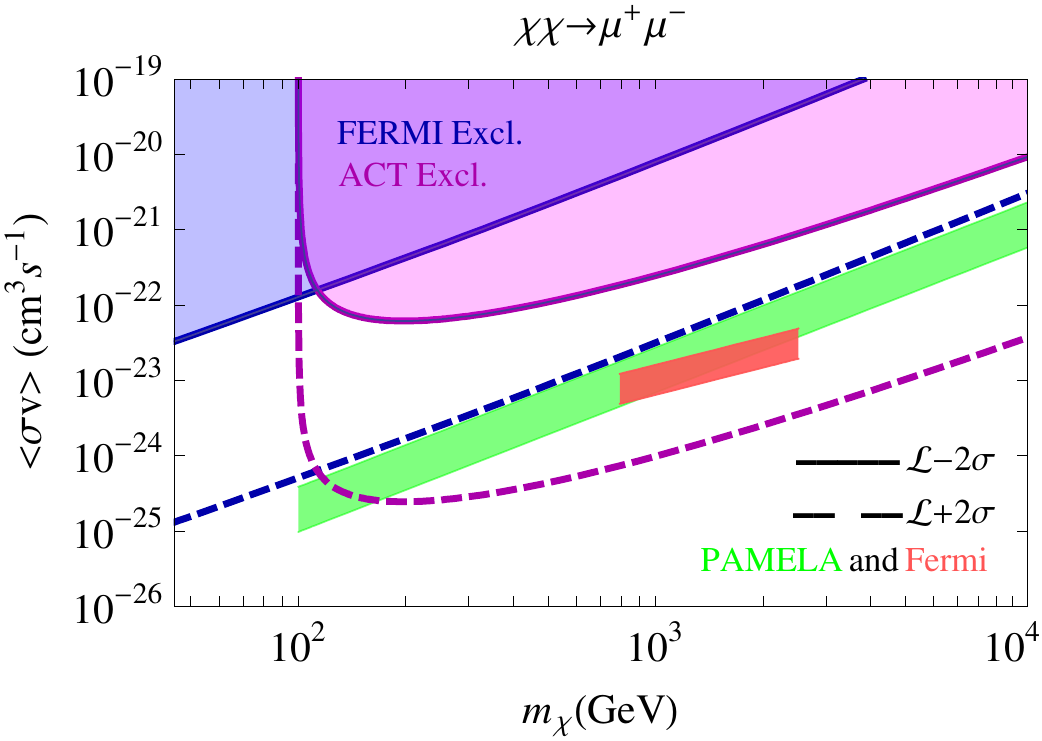}\\
\vskip 5mm
\includegraphics[width=.485\textwidth]{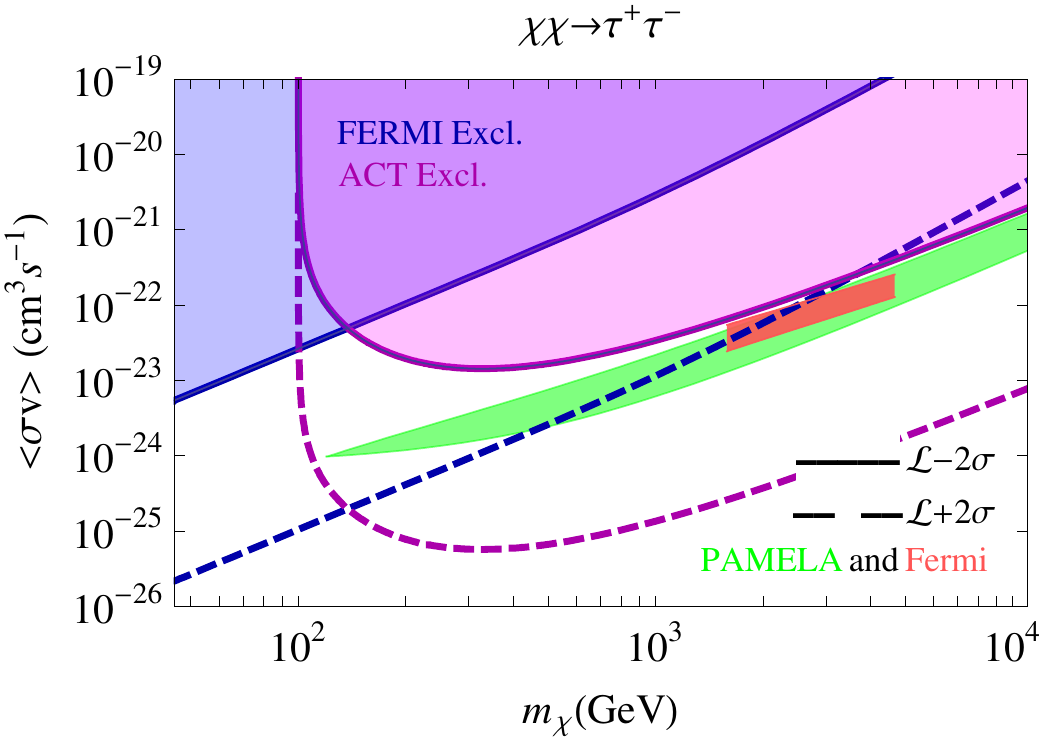}
\;
\includegraphics[width=.485\textwidth]{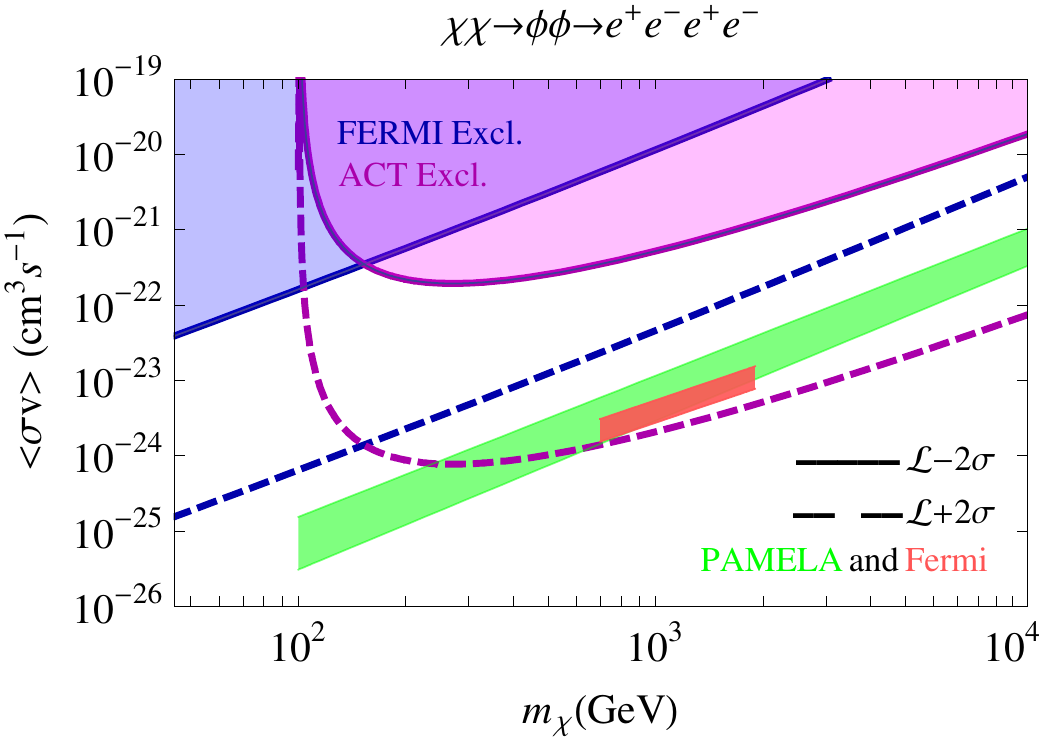}\\
\vskip 5mm
\includegraphics[width=.485\textwidth]{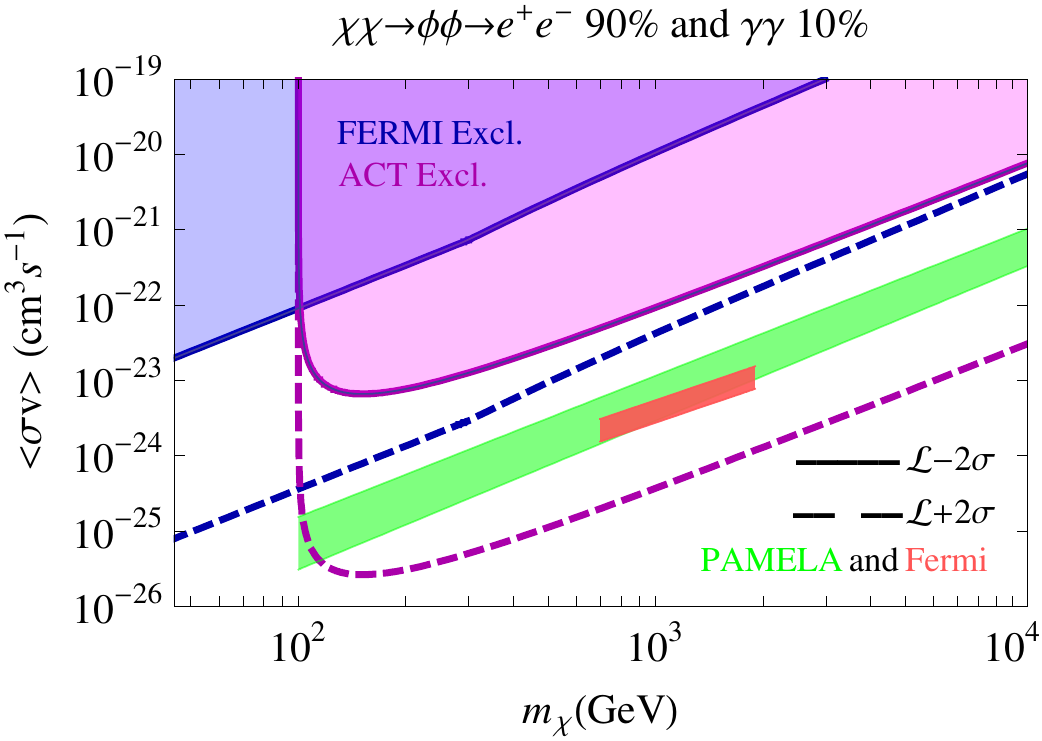}
\;
\includegraphics[width=.485\textwidth]{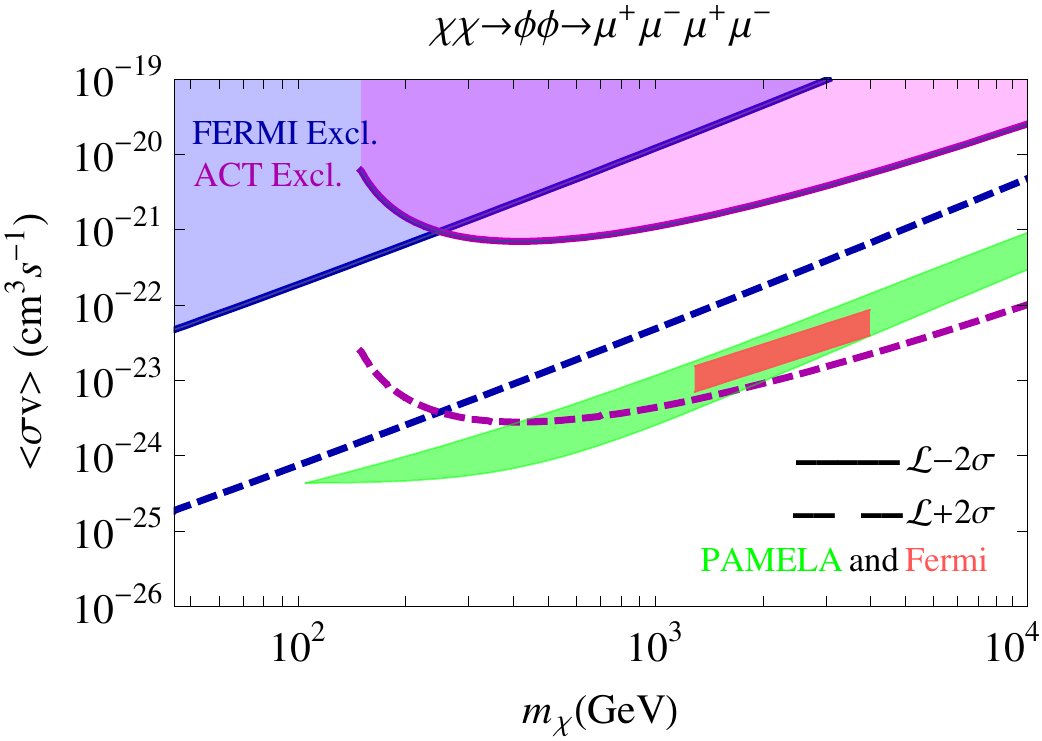}
\caption{Current exclusion regions from Fermi 9-month gamma-ray observations of Segue 1 (bounded below by the blue solid diagonal line) and MAGIC gamma-ray observations of Segue 1 (bounded below by the purple solid curved line).  
The exclusion regions use the conservative $2\sigma$ lower limit of $\mathcal{L}_{\rm{ann}}$ given in Table \ref{tab:SegLs} within $\theta = 0.25$ (Fermi) and $\theta=0.1$ (MAGIC).
The dashed blue and purple lines depict the respective 
cross-section bounds using the optimistic $2\sigma$ upper 
limits of $\mathcal{L}_{\rm{ann}}$.  
Green and orange regions indicate the best-fit regions to explain the PAMELA and Fermi
$e^+/e^-$ signals and are adapted from \cite{Meade:2009iu}  after rescaling from 
a local dark matter density of 0.3 GeV/cm$^3$ to 0.43 GeV/cm$^3$ \cite{Salucci:2010qr}.
}
\label{fig:gammabounds}
\end{center}
\end{figure*}

\begin{figure*}[t!h]
\begin{center}     
\includegraphics[width=.48\textwidth]{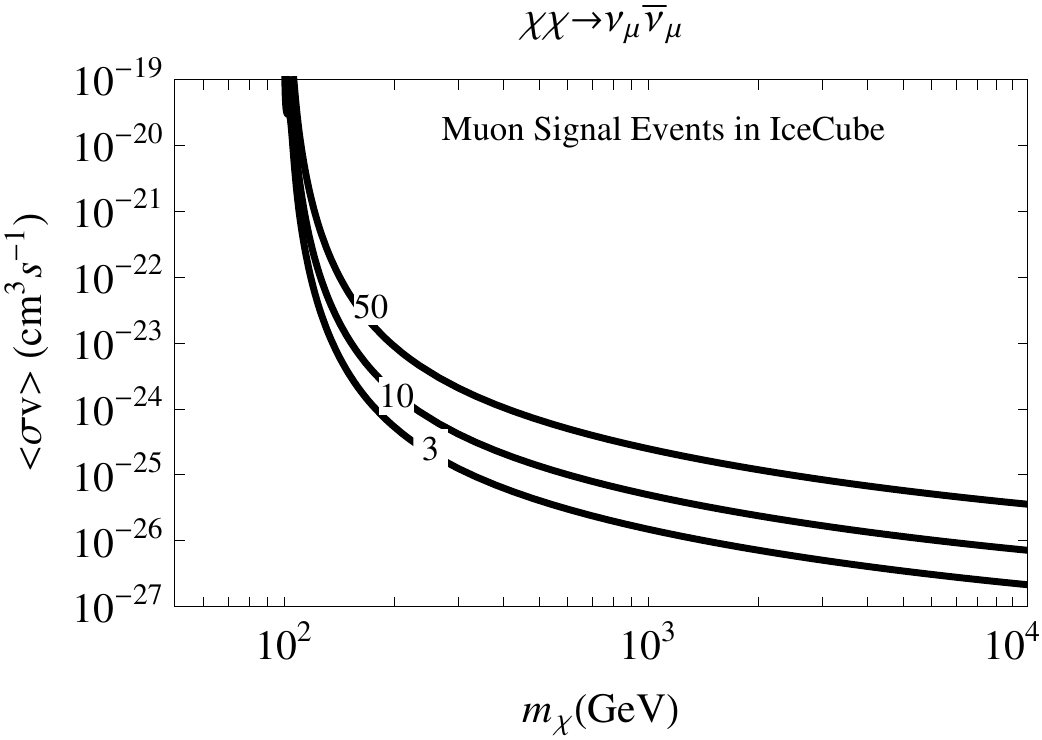}
\;
\includegraphics[width=.48\textwidth]{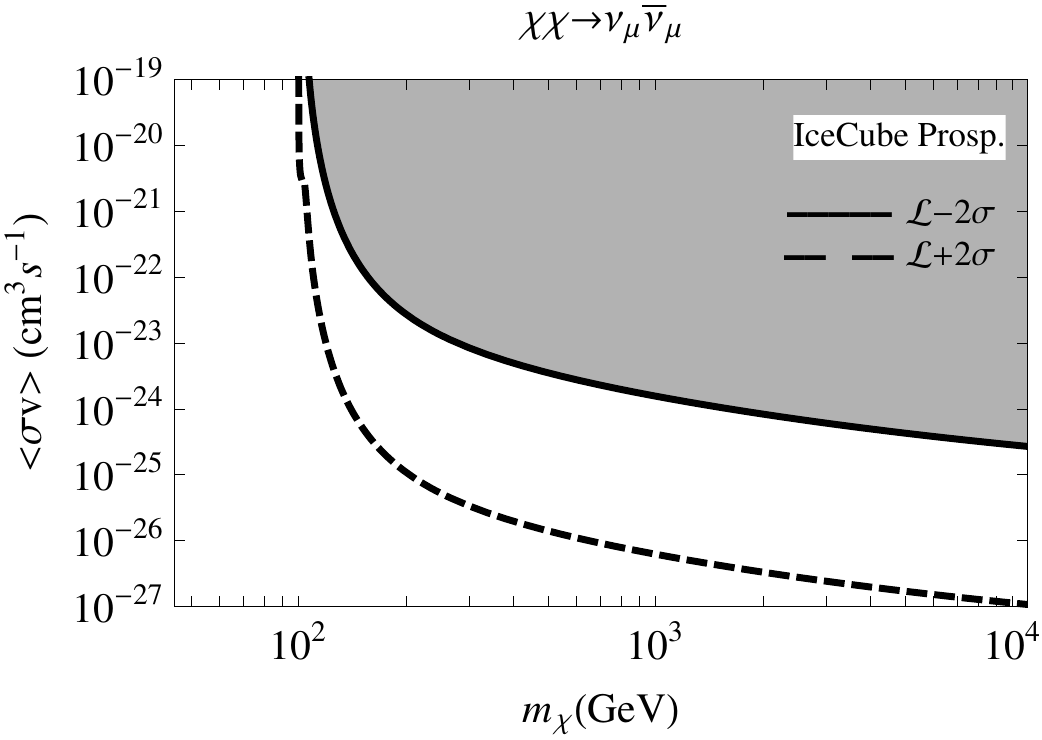}
\caption{{\bf Left:} Projected number of muon signal events coming from neutrinos originating from 
dark matter annihilation directly into $\nu_\mu\nu_\mu$ in Segue 1, after 10 years of observation 
with IceCube, and assuming the mean value of $\mathcal{L}_{\rm{ann}}^{\rm Segue 1}$ 
given in Eq.~(\ref{eq:Segue1}).
{\bf Right:} Prospective exclusion regions for IceCube 
10-year observations of Segue 1, assuming 
(signal muon events)/$\sqrt{\rm{signal+background~muon~events}} \ge 2$.
The muons observed by IceCube are assumed to come from muon 
neutrinos produced in dark matter annihilation.  Here the solid (dashed) line uses the 
$2\sigma$ lower (upper) limit on $\mathcal{L}_{\rm{ann}}$ in Eq.~(\ref{eq:Segue1}).} 
\label{fig:directneubounds}
\end{center}
\end{figure*}

Given $dN_{\nu_\mu, \bar\nu_\mu}/dE_\nu$ above, one obtains the differential neutrino flux, $d\Phi_{\nu_\mu,\bar\nu_\mu}/dE_\nu$, from Eq.~(\ref{equ:flux}).
The muon energy spectrum detected by IceCube in a time $T$ is given by
\begin{eqnarray}\label{eq:muonspectrum}
\frac{N_\mu}{dE_\mu} (E_\mu) & = & T \int_{E_{\mu}}^{m_\chi} dE_\nu
\frac{\rho_{\rm m}}{m_N}
\frac{d\Phi_{\nu_\mu,\bar\nu_\mu}}{dE_\nu}
\Big( \frac{d\sigma_\nu}{dE_\mu} +
\frac{d\sigma_{\bar\nu}}{dE_\mu}\Big) \nonumber \\
& & \times
R(E_\mu,E_{\rm th}) A_{\rm eff}(E_\mu).
\end{eqnarray}
Here the differential cross-section for a neutrino of energy $E_\nu$ to scatter off a
nucleon with mass $m_N$ is roughly given by \cite{Barger:1987,Winter:2000}
\begin{equation}
\frac{d\sigma_\nu}{dE_\mu} \simeq \frac{2 m_N G_F^2}{\pi} \Big(0.20 +
0.05 \frac{E_\mu^2}{E_\nu^2} \Big)
\end{equation}
while for an anti-neutrino the numbers 0.20 and 0.05 are interchanged.  For this 
expression,
we took the average of the scattering cross-sections off a neutron and off a proton, 
and we assumed $E_{\nu}\ll m_W^2/m_N \sim 7$ TeV, where $m_W$ is the $W$ boson mass.
(For dark matter masses of $\sim 10$ TeV that give rise to neutrinos of $\sim 10$ TeV, 
the constraints we obtain will thus be optimistic by a factor of a few.)  
Once produced at an energy $E_\mu$, a muon will loose energy according to
\begin{equation}
\frac{dE_\mu}{dr} \simeq - \rho_{\rm m}  (\alpha + \beta E_\mu),
\end{equation}
where $\alpha \simeq 2\times 10^{-3}$ GeV cm$^2$ g$^{-1}$, $\beta
\simeq 4.2 \times 10^{-6}$
cm$^2$ g$^{-1}$ \cite{Amsler:2008zzb}, and $\rho_{\rm m}$ is the Earth
matter density.
If we require that the muon be detected above an energy threshold of
$E_{\rm th}$, it will travel a distance
\begin{equation}
R(E_\mu,E_{\rm th}) \simeq \frac{1}{\rho_{\rm m}\beta}
\ln\Big( \frac{\alpha + \beta E_\mu}{\alpha + \beta E_{\rm th}}\Big).
\end{equation}
The muon must intersect IceCube's effective area, which for $E_\mu
\sim $10 TeV  is roughly 1 km$^2$,
but is much less for a lower energy muon.  We parameterize the
effective area, $A_{\rm eff}(E_\mu)$, for the
full IceCube detector following \cite{Ahrens2004,GonzalezGarcia:2005xw}.
To model IceCube's energy resolution, we also smear the muon energy spectrum given in 
Eq.~(\ref{eq:muonspectrum}) with
$\log_{10}(E_{\rm measured}/E_{\rm real})$ given by a gaussian
distribution centered on zero and with a standard deviation of 0.15 \cite{Resconi:2008fe}.
The total number of muons detected by IceCube is then given by
integrating this smeared energy spectrum 
over all muon energies from $E_{\rm th}$ up to $m_\chi$.

To calculate the muon background, we find $dN_{\nu_\mu, \bar\nu_\mu}/dE_\nu$ from atmospheric neutrinos 
using an average over zenith angle from \cite{Honda:2006qj}. We assume a constant pointing resolution of 
$1^\circ$ radius
down to neutrino energies of 100 GeV for IceCube \cite{Ahrens:2003ix}, which should be obtainable given 
Segue 1's zenith angle of $\sim 106^\circ$ with respect to the South Pole.

By fixing the ratio of signal muons to the square root of the signal plus background, 
$\langle\sigma v\rangle$ can be determined as a function of $m_\chi$.  

While the experimental properties quoted above refer specifically to IceCube, the inclusion of 
DeepCore \cite{Resconi:2008fe} in conjunction with IceCube will help in the reconstruction of muon events 
(ensuring a better pointing accuracy) and in the differentiation of background events.  

\section{Results}\label{sec:results}

\begin{figure*}[t!h]
\begin{center}     
\includegraphics[width=.48\textwidth]{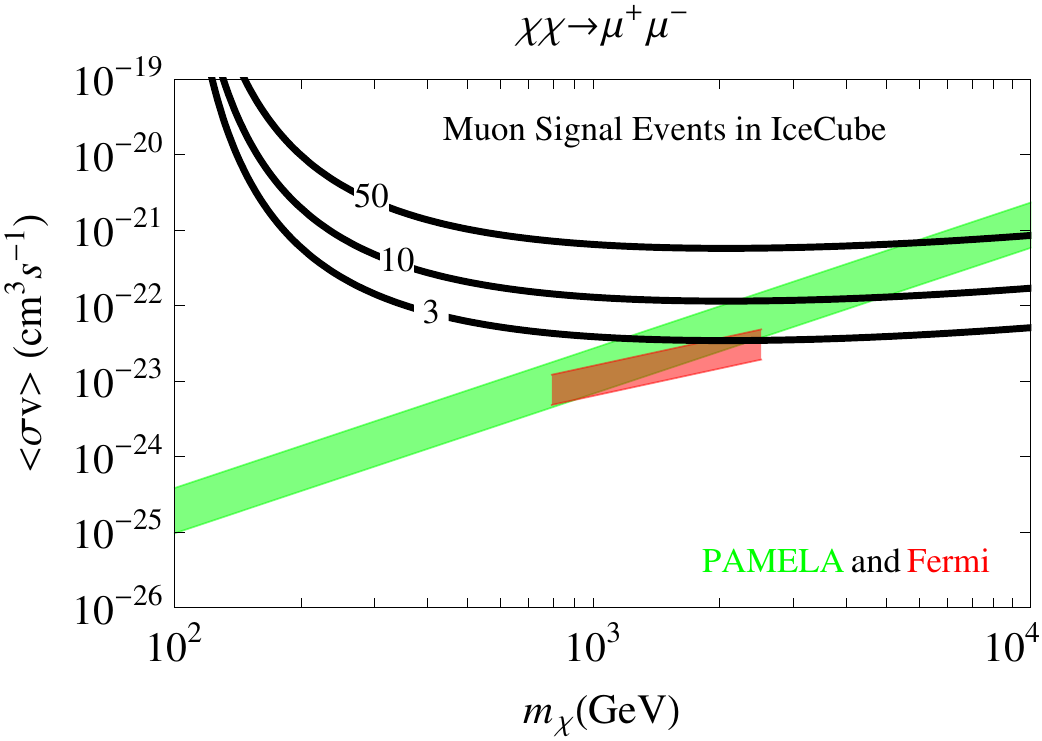}
\;
\includegraphics[width=.48\textwidth]{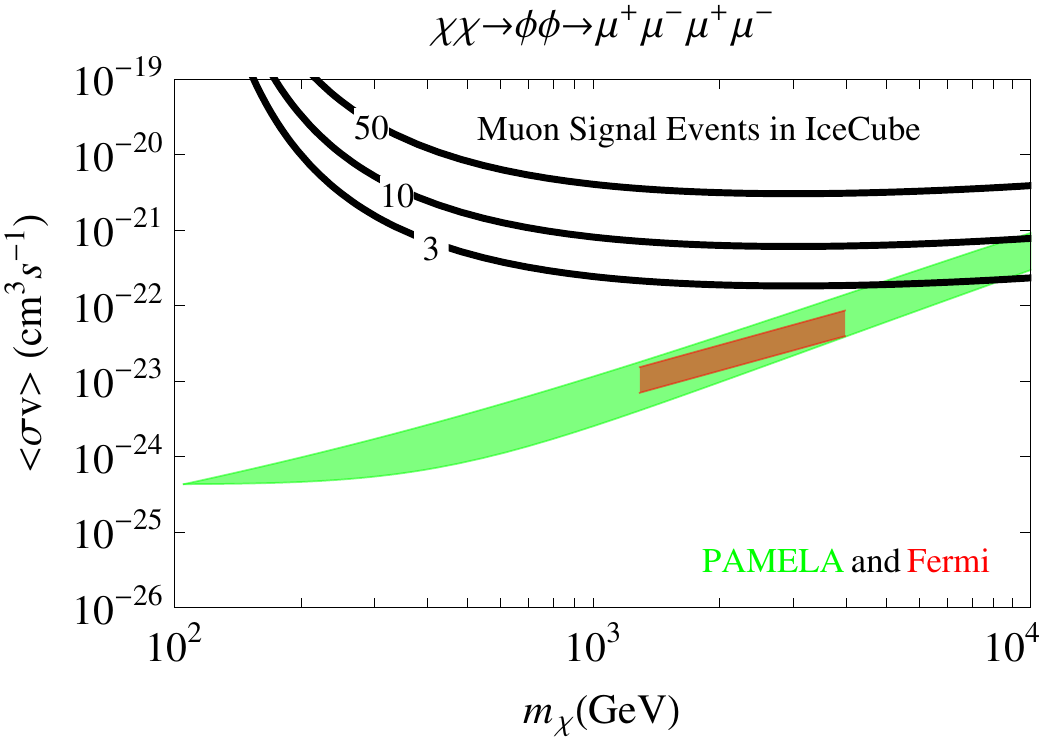}
\caption{Projected number of muon signal events coming from neutrinos originating from 
dark matter annihilation directly into muons (left) or into muons through an intermediate particle 
$\phi$ (right) in Segue 1, after 10 years of observation 
with IceCube, and assuming the mean value of $\mathcal{L}_{\rm{ann}}^{\rm Segue 1}$ 
given in Eq.~(\ref{eq:Segue1}).
Green and orange regions indicate the best-fit regions to explain the PAMELA and Fermi
$e^+/e^-$ signals and are adapted from \cite{Meade:2009iu} after rescaling from 
a local dark matter density of 0.3 GeV/cm$^3$ to 0.43 GeV/cm$^3$ \cite{Salucci:2010qr}.
Note that these projections do not assume any additional Sommerfeld enhancement in the 
dwarf galaxy compared to locally.  However, if present, such an enhancement could 
effectively shift up by an order of magnitude or more the PAMELA/Fermi regions in the right plot, 
increasing the expected number of signal events.
}
\label{fig:neuSignalEvents}
\end{center}
\end{figure*}

\subsection*{Current and Projected Observations}

Below we discuss current and projected experimental flux limits and resulting cross-section bounds.

{\bf Fermi} --- Fermi has reported gamma-ray constraints on 14 nearby dwarf galaxies from 11 months of data
in \cite{Abdo:2010ex}.  
For Segue 1, Fermi has only reported a flux limit of $4 \times 10^{-10} \,{\rm cm}^{-2}$ s$^{-1}$ 
for photons above 100 MeV and below 300 GeV
using 3 months of data \cite{Wang,Farnier}. 
They assumed a source spectrum of $\frac{dN_\gamma}{dE_\gamma} \sim \frac{1}{E_\gamma}$, 
which is very close to the FSR spectrum. 
 (Note that the photons from $\tau$ decays into $\pi^0$'s follow a different spectrum.)  
Since no signal was seen from Segue 1 in the 9-month data \cite{Murgia,Jeltema}, we conservatively 
take the flux limit from the 3-month data and divide by $\sqrt{3}$ to arrive at an approximate 
9-month flux limit ($2.3 \times 10^{-10} \,{\rm cm}^{-2}$ s$^{-1}$). 
(Fermi has not yet presented flux limits on a $1/E_\gamma$ spectrum using more than 
3 months of data.  The analysis in \cite{Scott:2009jn} uses 9 months of data to constrain 
the ``Constrained Minimal Supersymmetric Standard Model''.) 
This simple scaling expected from statistics is corroborated by Fermi's results for a $1/E_\gamma^2$ 
spectrum for which they did present results using 3 and 9 months of data: the flux limits for 
Segue 1 were $3.1 \times 10^{-9}  \,{\rm cm}^{-2}$ s$^{-1}$ \cite{Farnier} and 
$1.83\times 10^{-9}  \,{\rm cm}^{-2}$ s$^{-1}$ \cite{Jeltema,Murgia}, respectively.  
(It does, however, ignore any improvements that may have been made to the analysis 
in the meantime, which could further strengthen the limits.)
As expected, these limits are worse than the flux limit for a $1/E_\gamma$ signal spectrum, which has 
more photons at higher energies. 

{\bf ACTs} --- Segue 1 has been observed by MAGIC and a flux upper limit of  roughly $10^{-12}  \,{\rm cm}^{-2}$ s$^{-1}$ above 100 GeV has been reported \cite{magic}.  This limit should be viewed as an order of magnitude estimate as the exact limit will depend on the shape of the spectrum considered.  This limit is for observations of Segue 1 within 0.1 degrees of its center. 

{\bf IceCube} --- For IceCube, we show both the number of signal muon events expected and the 
cross-section bounds assuming (signal muon events)/$\sqrt{\rm signal+background~muon~events} > 2$.  We assume an energy threshold of 100 GeV, a pointing 
resolution of $1^\circ$ in radius that is constant with energy, and 10 years of observations.

\begin{figure*}[t!h]
\begin{center}     
\includegraphics[width=.48\textwidth]{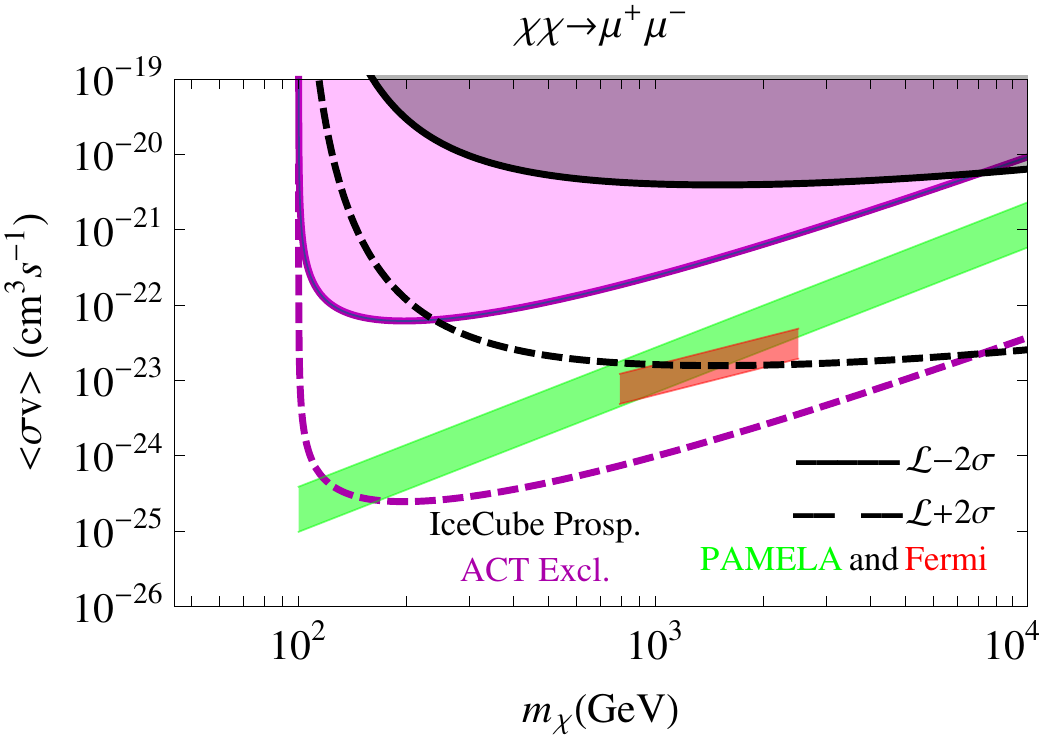}
\;
\includegraphics[width=.48\textwidth]{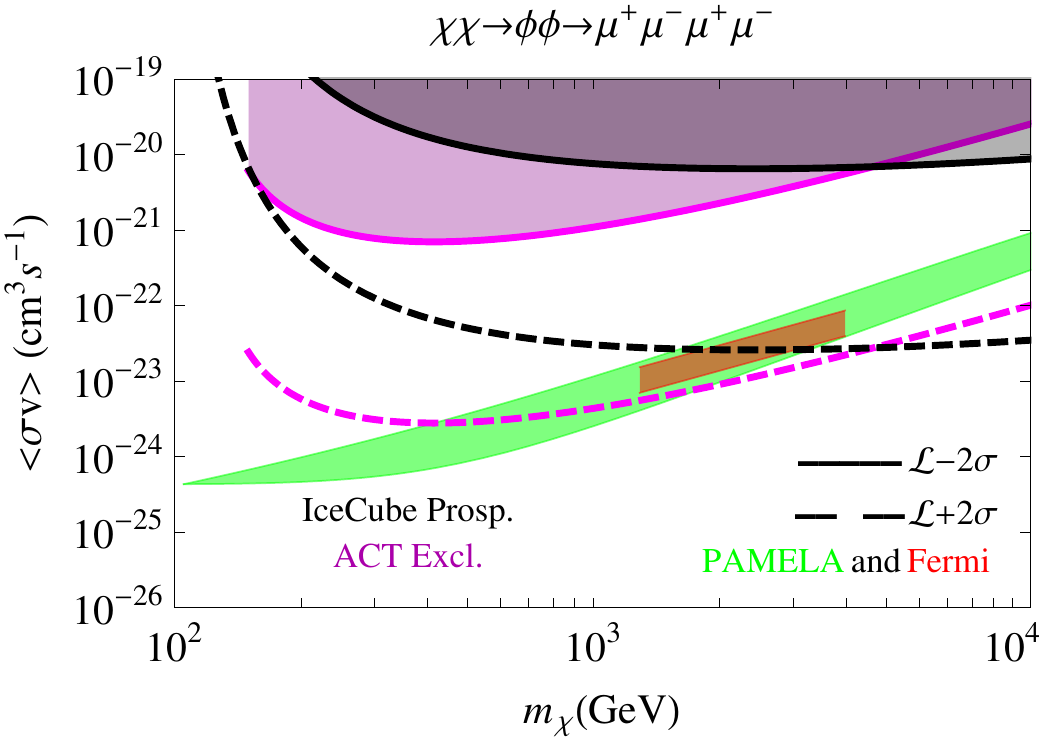}
\caption{Prospective exclusion regions for IceCube 
10-year observations of Segue 1.
The IceCube constraints assume 
(signal muon events)/$\sqrt{\rm{signal+background~muon~events}} \ge 2$.
The muons observed by IceCube are assumed to come from neutrinos produced by dark matter annihilation 
directly into muons (left) or into muons through an intermediate particle 
$\phi$ (right). The solid (dashed) line uses the 
$2\sigma$ lower (upper) limit on $\mathcal{L}_{\rm{ann}}$ in Eq.~(\ref{eq:Segue1}). 
Green and orange regions indicate the best-fit regions to explain the PAMELA and Fermi
$e^+/e^-$ signals and are adapted from \cite{Meade:2009iu} after rescaling from 
a local dark matter density of 0.3 GeV/cm$^3$ to 0.43 GeV/cm$^3$ \cite{Salucci:2010qr}.
} 
\label{fig:neubounds}
\end{center}
\end{figure*}

\subsection*{Cross-Section Bounds}

{\bf From Gamma-rays} --- 
In Figure~\ref{fig:gammabounds0}, we show the current upper bounds on the dark matter cross-section 
as a function of mass from Fermi and MAGIC observations of Segue 1.  We consider the channels 
$\chi\chi\to W^+W^-$, $b\bar{b}$, $t\bar{t}$, and $gg$.  
Figure~\ref{fig:gammabounds} shows the current upper bounds for the channels $\chi\chi \to e^+e^-$, $\chi\chi \to \mu^+\mu^-$, $\chi\chi \to \tau^+\tau$, 
$\chi\chi \to \phi\phi \to e^+e^-e^+e^-$, $\chi\chi \to \phi\phi \to \mu^+\mu^-\mu^+\mu^-$, 
and $\chi\chi \to \phi\phi$, with $\phi\to e^+e^-$ 90\% of the time, 
and $\phi\to \gamma\gamma$ 10\% of the time.  Here $\chi$ denotes the dark matter particle, 
while $\phi$ denotes an intermediate particle such as a gauge boson or a scalar.  For the 
case that $\phi$ goes to $\gamma\gamma$ 10\% of the time, $\phi$ is a light scalar that 
mixes with the Standard Model Higgs boson. 
The channels in Figure~\ref{fig:gammabounds} can give rise to the $e^+$ and $e^++e^-$ excesses 
observed by PAMELA and Fermi/HESS, as can various combinations of these channels 
with different branching ratios.
In all cases, we consider the mass of $\phi$ to be $m_\phi \lesssim \mathcal{O}$(GeV).  
This kinematically constrains the $\phi$ to decay to Standard Model particles that include a sizeable 
lepton fraction and avoids the overproduction of anti-protons.  This 
can also lead to Sommerfeld enhanced annihilation (see below).

In these figures, the solid lines
correspond to constraints using the mean $\mathcal{L}_{\rm{ann}}$ values \emph{minus} 
the $2\sigma$ errors as given in Table \ref{tab:SegLs} within $\theta=0.25$ (for Fermi) and $\theta=0.1$ (for MAGIC).  
Thus the solid lines show very conservative 
$95\%$ confidence level upper limits.  The dashed lines use the mean $\mathcal{L}_{\rm{ann}}$ values 
\emph{plus} the $2\sigma$ errors showing an optimistic reach.  
The green and orange regions indicate the best-fit regions to explain the PAMELA and Fermi
$e^+/e^-$ signals and are adapted from \cite{Meade:2009iu}.  Note, however, 
that we scaled down the regions in \cite{Meade:2009iu} from a local dark 
matter density of 0.3 GeV/cm$^3$ to 0.43 GeV/cm$^3$ \cite{Salucci:2010qr}.

Figures~\ref{fig:gammabounds0} and~\ref{fig:gammabounds} show that  Fermi has an advantage over ACTs for those dark matter channels that produce softer photons.   Fermi's much lower photon energy threshold compared to 
that of the ACTs ($\sim$100 MeV versus $\sim$100 GeV) also gives it an advantage in determining 
cross-section bounds for lower dark matter masses $\mathcal{O}$(100 GeV).    
ACTs have an advantage over Fermi in constraining models which produce harder photon spectra, 
assuming that the dark matter mass is above the ACT's threshold of around 100 GeV.  
This is because ACTs in general have larger collecting areas than Fermi ($\sim 10^4$ m$^2$ versus $\sim 1$ m$^2$) and thus have better statistics at high photon energies. 

In Figure \ref{fig:gammabounds0}, the results for the $\chi\chi\to W^+W^-$ channel show that the Wino 
LSP model can be probed by gamma-ray observations of Segue 1. 
This model, which can also explain the PAMELA data, predicts an annihilation cross-section of $\sim 2.5 \times 10^{-24} \,\mathrm{cm}^3$ s$^{-1}$ for $m_\chi \simeq 180$ GeV (indicated by the black dot) \cite{Kane:2009if}.  
While the observations disfavor 
this model (requiring a value for $\Lann$ less than its mean for consistency) they are unable to 
rule it out due to the uncertainty in $\Lann$. 
(Note that this model also predicts gamma-ray lines, and limits on such lines 
disfavor it \cite{Abdo:2010nc}).  
A more precise determination by the Fermi collaboration of the flux limit, using the photon spectrum 
for dark matter annihilation into $W^+W^-$, may set a stricter limit.

In Figure~\ref{fig:gammabounds}, we see that ACTs give stricter constraints than Fermi 
on dark matter models favored by the PAMELA and/or Fermi/HESS cosmic-ray electron and positron 
data.  
In many cases, the mean value for $\Lann$ probes the relevant cross-section and mass 
region, although taking the conservative value for $\Lann$ only constrains the 
$\chi\chi\to\tau^+\tau^-$ channel.  
Below we discuss in detail how Sommerfeld enhanced annihilation can be probed with Segue 1, 
which can lead to stronger constraints in the case of dark matter annihilation to $\phi$'s.

\vspace{2mm}

{\bf From Neutrinos} --- 
In Figure~\ref{fig:directneubounds} (left), we show the 
number of signal muon events expected as a function of  
$\langle\sigma v\rangle$ and $m_\chi$, assuming the mean $\mathcal{L}_{\rm ann}$ value for 
Segue 1 in Eq.~(\ref{eq:Segue1}), for the channel $\chi\chi\to \nu_\mu \bar{\nu}_\mu$, while 
Figure~\ref{fig:neuSignalEvents} shows the same for the channels 
$\chi\chi \to \mu^+\mu^-$ (left) and
$\chi\chi \to \phi\phi \to \mu^+\mu^-\mu^+\mu^-$ (right).  

Figure~\ref{fig:directneubounds} gives an indication of how well IceCube can probe 
the ``best-case'' scenario of dark matter annihilations directly into neutrinos.   
The right plot in this figure shows the expected neutrino exclusion limits from 
IceCube observations of Segue 1 for this channel.  

Figure~\ref{fig:neuSignalEvents} shows that if dark matter with $m_\chi > $1 TeV 
explains the PAMELA and/or Fermi 
excesses, one could expect a few muon signal events from Segue 1 for the first of these channels, but 
perhaps only one for the annihilation channel through an intermediary $\phi$.  
Note that the PAMELA/Fermi best fit regions assume a local dark matter density of 
0.43 GeV/cm$^3$ \cite{Salucci:2010qr} --- if the local dark matter density is lower, then a few more 
neutrinos could be expected.
Moreover, if the $\mathcal{L}_{\rm ann}$ value is larger than the mean, then also a few more could be 
expected.  
The number of expected neutrinos could also be larger in the case of  Sommerfeld enhanced models, 
where the low velocity dispersion of the dwarf galaxy may lead to a larger dark matter 
annihilation cross-section in the dwarf galaxy compared to locally (see below).
Figure~\ref{fig:neubounds} shows the expected neutrino exclusion limits from IceCube observations of 
Segue 1 for the same channels.  
For both annihilation channels, the ACT prospects are better than those from IceCube for low dark 
matter masses; for high dark matter masses (above the preferred PAMELA/Fermi regions) both are comparable.  
However, the observing time for ACTs is of order a few days whereas for IceCube it is years.  

We note that neutrinos from dwarf galaxies are less sensitive probes of dark matter annihilation 
to charged leptons than are observations of neutrinos from the Galactic center 
region \cite{Spolyar:2009kx}.  
However, observations of different dwarf galaxies may be stacked, increasing the sensitivity by 
a factor of a few, and any resulting neutrino signal together with a corroborating gamma-ray 
signal would be very strong evidence for dark matter annihilation. 

\subsection*{Constraining Sommerfeld enhanced models}

\begin{figure*}[t!h]
\begin{center}     
\includegraphics[width=.48\textwidth]{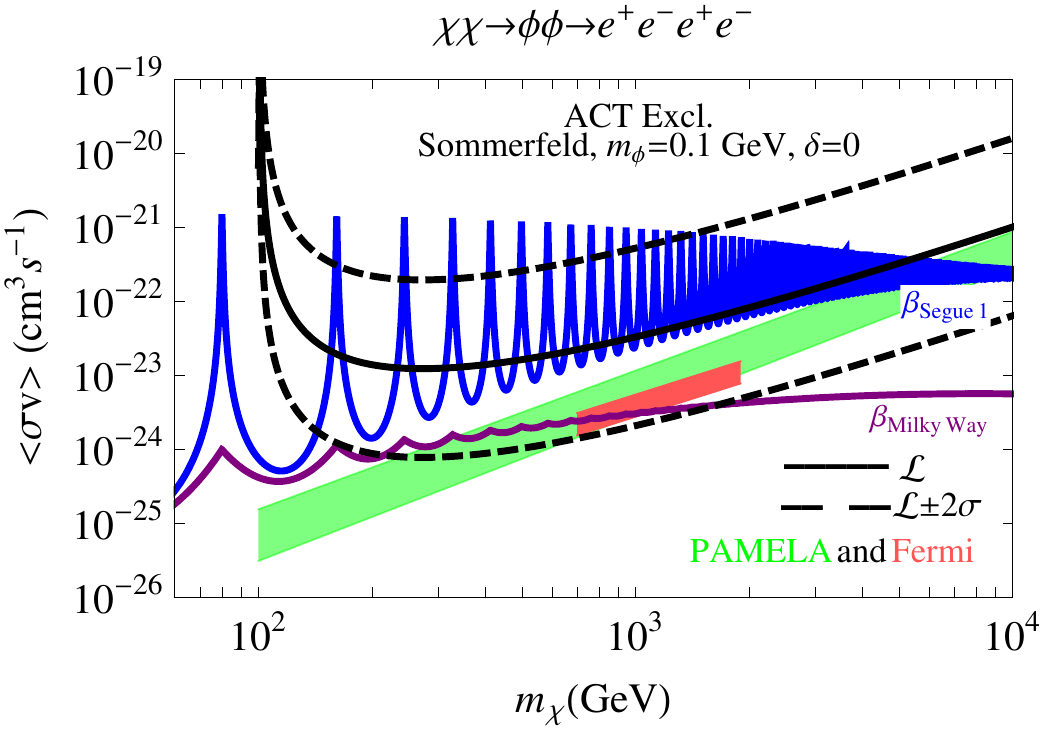}
\;
\includegraphics[width=.48\textwidth]{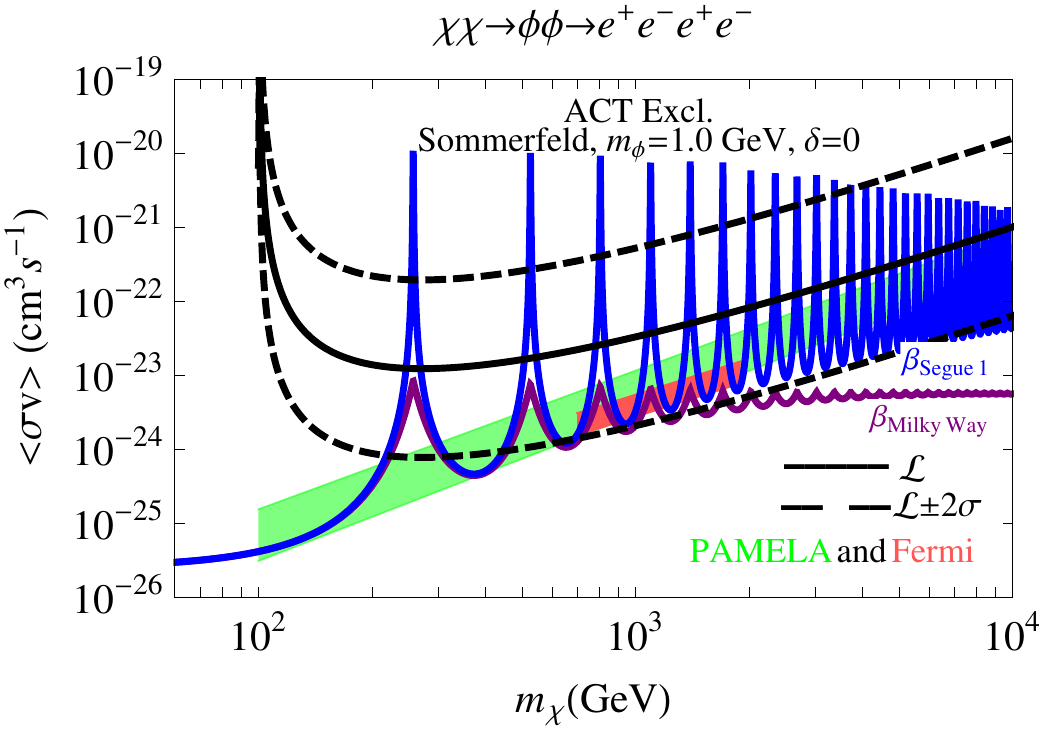} \\
\vskip 5mm
\includegraphics[width=.48\textwidth]{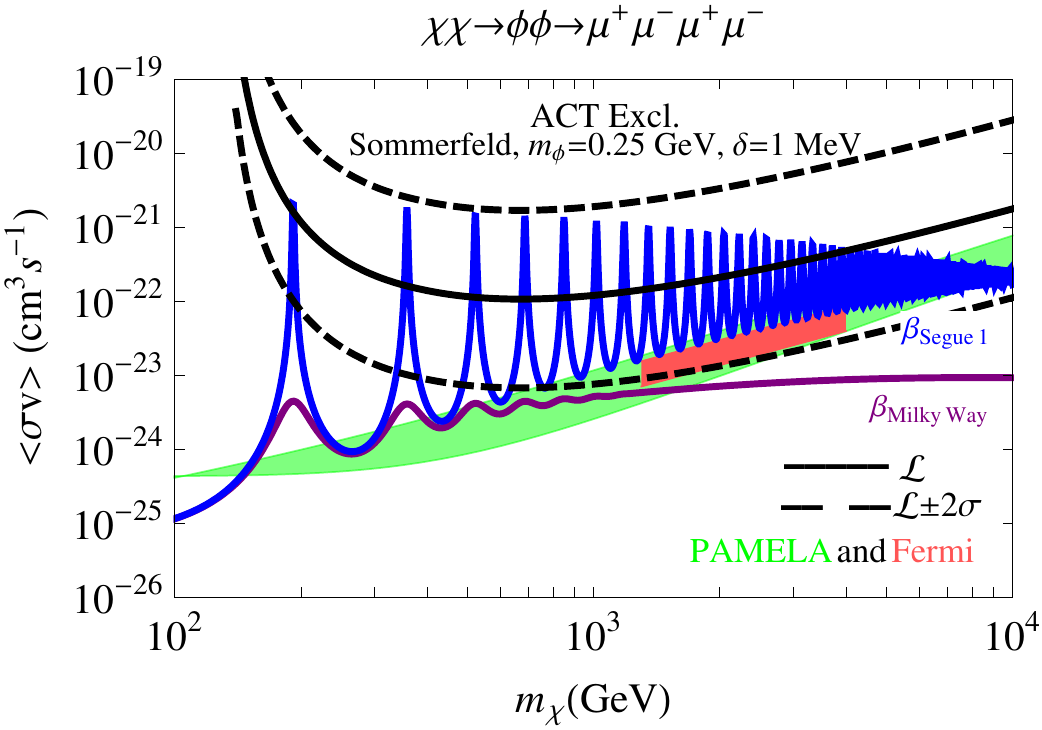}
\;
\includegraphics[width=.48\textwidth]{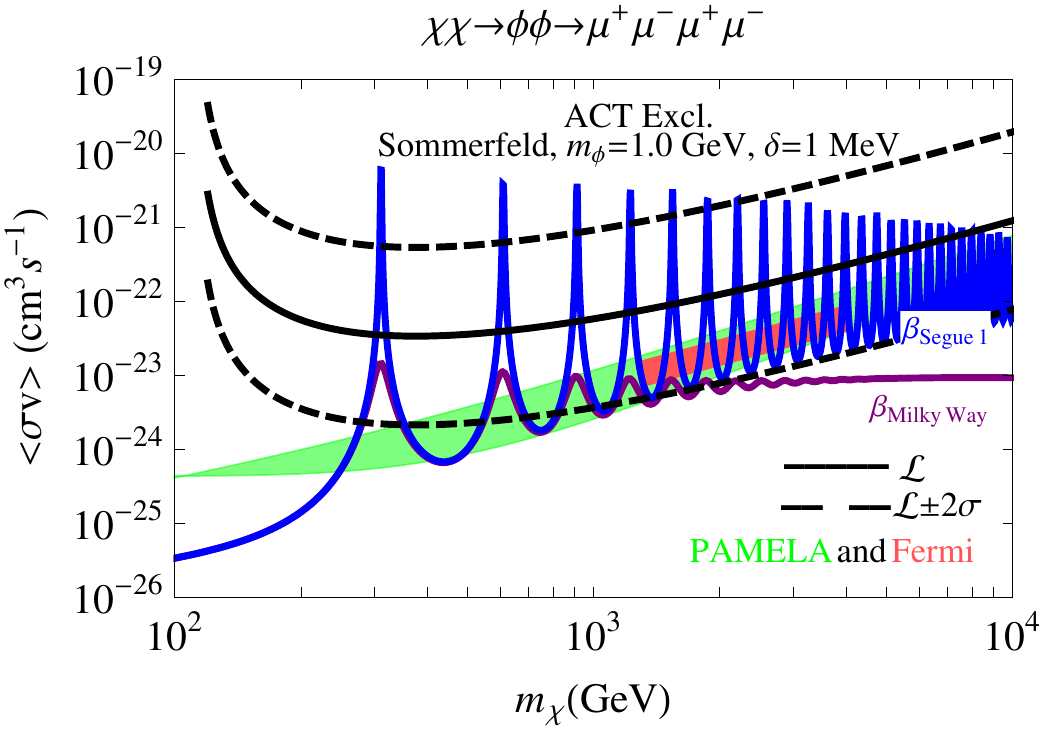}
\caption{The Sommerfeld enhanced annihilation cross-section as a function of dark matter 
mass $m_\chi$ for a mediator mass of $m_\phi=0.1$ GeV and dark matter mass 
splitting $\delta=0$ (upper left), 
$m_\phi=1$ GeV and $\delta=0$ (upper right), $m_\phi=0.25$ GeV and $\delta=1$ MeV 
(lower left), and $m_\phi=1.0$ GeV and $\delta=1$ MeV 
(lower right) assuming the dark matter velocity dispersion of the Milky-Way halo 
(lower purple line) or Segue 1 (upper blue line).  
Here $\delta=0$ refers to dark matter without excited states, while $\delta\ne 0$ refers to 
the splitting between the dark matter ground state and an excited state.  
The coupling $\alpha_D$ between the dark matter and $\phi$ is a function of 
$m_\chi$ and has been set to the value that gives the correct thermal relic abundance.  
Note that $\beta \equiv v/c$.  
Limits from MAGIC observations of Segue 1 are shown in black solid or dashed lines, where 
the solid line uses the mean value of $\mathcal{L}_{\rm ann}$ and the upper (lower) dashed line 
uses the $2\sigma$ lower (upper) bound on $\mathcal{L}_{\rm ann}$ for Segue 1 (see 
Eq.~(\ref{eq:Segue1})).  
Green and orange regions indicate the best-fit regions to explain the PAMELA and Fermi
$e^+/e^-$ signals for $\phi\to e^+e^-$ (upper plots) and for $\phi\to\mu^+\mu^-$ (lower plots)
and are adapted from \cite{Meade:2009iu} after rescaling from 
a local dark matter density of 0.3 GeV/cm$^3$ to 0.43 GeV/cm$^3$ \cite{Salucci:2010qr}.
Note that the purple lines intersect, or at least come close to, the PAMELA/Fermi preferred regions.  
This figure shows that a wide range of values for $m_\phi$ can be probed and that, in particular, 
$m_\phi \sim 0.1$ GeV is slightly disfavored by current constraints (upper left plot).}
\label{fig:Sommerfeld}
\end{center}
\end{figure*}

In order for annihilating dark matter to explain the PAMELA and/or Fermi/HESS $e^-/e^+$
excesses, the annihilation cross-section must be 
$\mathcal{O}(100-1000)$ times larger than the thermal WIMP freeze-out cross-section, which we take to 
be 
\be\label{eq:freeze-out}
\langle\sigma v_{\rm rel}\rangle_{\rm f.o.} \simeq 2.5 \times 10^{-26} \,{\rm cm}^3 {\rm s}^{-1}.
\ee
If the WIMP was produced non-thermally in the early Universe (for example, from the 
decay of a heavier particle), it can naturally have a large annihilation 
cross-section \cite{Kane:2009if,Kamionkowski:1990,Moroi:1999zb,Giudice:2000ex}.   
However, even if the WIMP was in thermal equilibrium in the early Universe, several 
mechanisms allow the cross-section to be large today while being 
much smaller at freeze-out \cite{Ibe:2008ye,Sommerfeld,Hisano:2003ec,Cirelli:2008pk,Arkani-Hamed2008, Pospelov:2008jd,MarchRussell:2008yu,MarchRussell:2008tu}.  
The Sommerfeld enhancement 
\cite{Sommerfeld,Hisano:2003ec,Cirelli:2008pk,Arkani-Hamed2008, Pospelov:2008jd,MarchRussell:2008yu,MarchRussell:2008tu} 
increases the cross-section at low dark matter velocities, $v$,  
so that the cross-section can be much larger today in the Milky-Way halo, where 
$v \sim 10^{-3}$, than during freeze-out, where $v\sim 0.3$.  
Since the dark matter velocity in dwarfs is $v \sim 10^{-4}$, there may be an additional order of 
magnitude
enhancement in the annihilation cross-section in dwarfs 
(see also \cite{Robertson:2009bh,Kuhlen:2009jv,Kuhlen:2009kx}).  
This additional enhancement is not guaranteed, however, since the Sommerfeld 
effect may saturate for very low $v$. 
(In this paper, we do not consider any additional enhancement from 
the even colder substructure within the dwarf galaxy, 
but see for example \cite{Bovy:2009zs}.)

The Sommerfeld enhancement occurs when the dark matter particle $\chi$ couples to a new 
mediator $\phi$, with mass $m_\phi\ll m_\chi$, that produces an attractive potential between 
the dark matter particles.
The $\phi$ in turn couples to Standard Model matter --- in particular, we imagine that $\phi$ can decay 
to $e^+e^-$ or $\mu^+\mu^-$, so that dark matter annihilation $\chi\chi\to\phi\phi\to e^+e^-e^+e^-$ or 
$\chi\chi\to\phi\phi\to \mu^+ \mu^- \mu^+ \mu^-$, or a combination, 
can produce the cosmic-ray excesses (in many models, and depending on $m_\phi$, 
$\phi$ decays will also produce light hadrons).  
In this section, we show that gamma-ray observations of Segue 1 disfavor low masses for $\phi$,
$m_\phi \sim \mathcal{O}$(0.1 GeV), assuming the mean value for $\Lann$, 
although the uncertainty in $\mathcal{L}_{\rm ann}$ does not allow 
one to conclusively rule out such low masses.  

To calculate the Sommerfeld enhancement, we assume that the dark matter couples to a 
mediator $\phi$ with coupling strength $\lambda$, and we consider two cases.  
The first case assumes that the dark matter interaction with $\phi$ is purely \emph{elastic}, while the 
second case assumes that the interaction is purely \emph{inelastic}.  The latter case occurs in the 
presence of an additional excited dark matter state $\chi^*$ so that the interaction with $\phi$ is only 
off-diagonal (i.e.~$\chi$ couples only inelastically to $\chi^*\phi$, and not elastically to 
$\chi\phi$).  We denote the mass splitting between $\chi$ and $\chi^*$ by $\delta$.  
The Sommerfeld enhancement in the presence of an excited state has been 
investigated in detail in \cite{Slatyer:2009vg}, where it was found that generically it predicts larger 
enhancements compared to the elastic case.  
Excited states occur in many models, including the ``inelastic dark-matter'' 
\cite{TuckerSmith:2001hy} and ``exciting dark matter'' models \cite{Finkbeiner:2007kk}. 

For the elastic case, we model the interaction with an attractive Yukawa potential of the form 
(following \cite{Arkani-Hamed2008})
\begin{equation}\label{eq:Yukawa}
V(r) = - \frac{\alpha_D}{r} e^{-m_\phi r},
\end{equation}
where $\alpha_D = \lambda^2 / 4\pi$. 
The wave-function of two annihilating dark matter particles in the non-relativistic limit 
is found by solving the Schr\"odinger equation,
\begin{equation}\label{eq:Schroedi}
\psi''(r) - m_\chi V(r) \psi(r) + v^2 m_\chi^2 \psi(r) = 0, 
\end{equation}
with the boundary condition $\psi'(\infty) = i v m_\chi \psi(\infty)$ and with 
$\psi(0)$ arbitrary.  
The total annihilation cross-section is then given by
\bea\label{eq:Sommerfeld_elastic}
(\sigma v_{\rm rel}) & = & S_0 (\sigma v_{\rm rel})_0 \nonumber \\
& \equiv & \left|\frac{\psi(\infty)}{\psi(0)}\right|^2 (\sigma v_{\rm rel})_0,
\eea
where $v=v_{\rm rel}/2$ is the dark matter particle's velocity in the center-of-mass frame, 
$S_0$ denotes the Sommerfeld enhancement in the elastic case, and 
\be\label{eq:cross-section}
(\sigma v_{\rm rel})_0 = \f{\pi \alpha_D^2}{m_\chi^2}
\ee
is the cross-section for $\chi\chi\to \phi\phi$ annihilation.  
This cross-section is modified if there are additional states into which $\chi$ can annihilate, 
but we will not consider this possibility (however see \cite{Feng:2010zp,Finkbeiner:2010}).  
In the limit $v \ll 1$ and $m_\phi \ll m_\chi$, Eq.~(\ref{eq:Schroedi}) may 
be solved exactly to give
\be\label{eq:Sommerfeld_simple}
S_0 = \frac{\alpha_D \pi}{v} \Big(1-e^{-\alpha_D \pi / v}\Big),
\ee 
which clearly shows the $1/v$ dependence of the cross-section.  
Away from this limit, the solution to equation (\ref{eq:Schroedi}) must be numerically 
integrated; it has resonances and saturates when $v \sim m_\phi/m_\chi$.  
Note that for $\alpha_D \ll v$, Eq.~(\ref{eq:Sommerfeld_simple}) can be expanded 
to give 
\be\label{eq:Sommerfeld_verysimple}
S_0 \simeq 1 + \f{1}{2}\f{\pi \alpha_D}{v} + \f{1}{12}\Big(\f{\pi \alpha_D}{v}\Big)^2 .
\ee 

A useful analytical formula \cite{MarchRussell:2008tu,Chen2007} that closely 
matches the numerical solution can be obtained by using 
the Hulth\'en potential $V(r) = -\alpha_D m_\phi e^{-r m_\phi}/(1-e^{-r m_\phi})$, which has 
similar behavior to the Yukawa potential for $r\to 0$ and $r\to \infty$.  The solution is given by 
\be\label{eq;Sommerfeld_delta0}
S_0 (v) = \Big(\f{\pi}{\epsilon_v}\Big) \, 
\f{\sinh\left(\f{12 \epsilon_v}{\pi \epsilon_\phi}\right)}{
\cosh\left(\f{12 \epsilon_v}{\pi \epsilon_\phi}\right) - 
\cos\left(2\pi \sqrt{\f{6}{\pi^2\epsilon_\phi} - \f{36 \epsilon_v^2}{\pi^4 \epsilon_\phi^2}}\right)},
\ee
where 
\be
\epsilon_v = \f{v}{\alpha_D c},\;\;\; \epsilon_\phi = \f{m_\phi}{\alpha_D m_\chi}.
\ee
We will use this analytic formula in the calculation of the Sommerfeld enhancement in the elastic 
case as described below. 

In order to calculate the Sommerfeld enhancement in the Milky-Way halo and in Segue 1 today, 
we proceed as follows (see also \cite{Feng:2009hw,Feng:2010zp,Finkbeiner:2010}).  
First we use Eqs.~(\ref{eq:cross-section}) and (\ref{eq:Sommerfeld_verysimple}) to set 
the annihilation cross-section in Eq.~(\ref{eq:Sommerfeld_elastic}) 
equal to the cross-section that gives the correct thermal relic abundance during freeze-out, 
Eq.~(\ref{eq:freeze-out}).  This allows us to capture the Sommerfeld effect during 
freeze-out, where, although much less important than in the Milky-Way and dwarf halo now, 
it  can still enhance the dark matter annihilation cross-section by an $\mathcal{O}(1)$ amount 
\cite{Yuan:2009bb,Dent:2009bv,Zavala:2009mi,Feng:2010zp,Finkbeiner:2010}. 
For a given $m_\chi$, this allows us to calculate $\alpha_D$.
The precise value of $m_\phi$ is not very important, and the use of the simple formula 
Eq.~(\ref{eq:Sommerfeld_verysimple}) as opposed to its more complicated variations 
is sufficient.  
Moreover, we use $v \sim 0.1$ (as opposed to the canonical value $v \sim 0.3$) 
for the velocity of the dark matter during freeze-out and have checked 
that the resulting value of $\alpha_D$ as a function of $m_\chi$ agrees to within about 10\% 
with a detailed calculation of the thermal relic abundance in \cite{Feng:2010zp}.  
The resulting value of $\alpha_D$ may now be used to calculate the Sommerfeld enhancement 
in the Milky-Way and Segue 1 halo. 
In particular, we take the dark matter distribution to be 
thermal so that the thermally averaged Sommerfeld enhanced annihilation cross-section is 
given by \cite{Feng:2010zp}
\bea\label{eq:Sommerfeld_averaged}
\langle \sigma v_{\rm rel} \rangle & \simeq & (\sigma v_{\rm rel})_0 \, \bar{S}_0,
\eea
with  
\bea
\bar{S}_0 & \equiv &  \f{(2/v_0^2)^{3/2}}{2\sqrt{\pi}N}\int_0^{v_{\rm max}}\, 
		S_0\Big(\f{v_{\rm rel}}{2}\Big) \, e^{-\f{v_{\rm rel}^2}{2v_0^2}} dv_{\rm rel},
\eea
where $N = {\rm{erf}}(z/\sqrt{2}) - \sqrt{2/\pi}\, z\, e^{-z^2/2}$ with $z\equiv v_{\rm esc}/v_0$, 
$v_{\rm esc}$ is the dark matter escape velocity, and $v_{\rm max}$ is the maximum 
relative dark matter velocity in the laboratory frame, which we will simply set equal to 
$v_{\rm esc}$ .  
Here we use the analytic formula for $S_0$ given 
in Eq.~(\ref{eq;Sommerfeld_delta0}), and take for the Milky-Way 
\cite{Xue:2008se}
\be
v_0 \simeq 210~{\rm km/s}, \;\;\; v_{\rm esc} \simeq v_{\rm max} \simeq 525~{\rm km/s} 
\ee
(these values are close to other values in the literature, e.g.~\cite{Reid:2009nr,Smith:2006ym}, 
and allow us to compare our results to those in \cite{Feng:2010zp}).  
For Segue 1, the average one-dimensional dark matter velocity dispersion within a 
$0.25^\circ$ radius from the center is $4.5 \pm 1.8$ km/s.  
This was obtained using Eq.~(\ref{eq:sigma_dm}).  
The relevant parameters we use for Segue 1 are 
\be
v_0 \simeq \sqrt{2}\times 4.5~{\rm km/s} \simeq 6.4~{\rm km/s},
\; v_{\rm esc} \sim v_{\rm max} \sim 10~{\rm km/s}. 
\ee
We note that small variations in the value of $v_{\rm esc}$ and $v_{\rm max}$ for Segue 1 
do not qualitatively impact the results.  

In Figure \ref{fig:Sommerfeld} (upper two figures), we show the resulting Sommerfeld enhanced 
annihilation cross-section as a function of $m_\chi$ for $m_\phi = 0.1$ GeV (upper left) 
and $m_\phi = 1$ GeV (upper right) for dark matter annihilating in the Milky-Way halo (purple line)
and in Segue 1 (blue line).  We superimpose these on the PAMELA/Fermi preferred region adapted 
from \cite{Meade:2009iu}, for the case that $\phi$ decays exclusively to $e^+e^-$, 
after rescaling their values from a local dark matter density of 0.3 GeV/cm$^3$ to 
0.43 GeV/cm$^3$ \cite{Salucci:2010qr}\footnote{We note that the cosmic-ray spectrum 
obtained from $\phi$'s decaying exclusively to $e^+e^-$ gives a slightly worse fit to the PAMELA/Fermi 
spectra than if one, for example, assumes additional hidden sector showering,  
or $\phi$ decays to electrons and muons, or only to muons \cite{Meade:2009iu}.  We refer the reader to 
\cite{Finkbeiner:2010} for a detailed discussion and benchmark points consistent with the 
PAMELA/Fermi spectra and preferred regions in the cross-section and mass plane.}.  
Note that the purple line does intersect the PAMELA/Fermi preferred region, as it must if 
it is to explain this data.  
The gamma-ray constraints from MAGIC observations of 
Segue 1 are shown in black lines, with the solid line using 
the mean value of $\Lann$, while the upper dashed shows the conservative limit 
with $\Lann-2\sigma$, and the lower dashed line shows the optimistic limit with $\Lann+2\sigma$.  
We see that the additional Sommerfeld enhancement expected in Segue 1 (indicated by 
the blue line) allows us to constrain this model. 
In particular, low values of $m_\phi$ are disfavored assuming the mean value for 
$\Lann$, although they cannot be ruled out due to the uncertainty in $\Lann$.  
Since the Sommerfeld enhancement saturates when $v\lesssim m_\phi/m_\chi$, 
larger $m_\phi$ are less constrained.  However, even for larger $m_\phi \sim \mathcal{O}$(1 GeV) 
a sizeable fraction of the parameter space lies near a resonance region, and is constrained.  

For the elastic case and for $\phi$ decaying exclusively to $\mu^+\mu^-$, the $\alpha_D$ required 
by thermal freeze-out is too small to give a large enough Sommerfeld enhanced annihilation 
cross-section to explain the PAMELA/Fermi regions \cite{Feng:2010zp}.
Inelastic interactions increase the cross-section and alleviate the 
discrepancy \cite{Slatyer:2009vg,Finkbeiner:2010}.  
In Figure \ref{fig:Sommerfeld} (lower two plots), we show 
the resulting Sommerfeld enhanced annihilation cross-section as a function of $m_\chi$ for 
$\delta = 1$ MeV for $m_\phi = 0.25$ GeV (lower left) and $m_\phi = 1$ GeV (lower right) for 
dark matter annihilating in the Milky-Way halo (purple line) and in Segue 1 (blue line).  
We review the details of the Sommerfeld enhancement in the presence of excited 
states in Appendix \ref{app:inelastic} (note that we ignore any complications in the 
freeze-out calculation of having additional excited states).
For the case in which $\phi$ decays exclusively to $\mu^+\mu^-$, we superimpose these 
on the PAMELA/Fermi preferred region adapted from \cite{Meade:2009iu}, again rescaling their 
results to a local dark matter density of 0.43 GeV/cm$^3$ (note, however, that the Sommerfeld 
cross-sections are independent of the $\phi$ decay modes).
The gamma-ray constraints from MAGIC observations of Segue 1 are again shown in black 
lines, and we find that for both small and larger values of $m_\phi$, they constrain part of the relevant 
parameter space.  

The purple lines in the two lower plots of Figure \ref{fig:Sommerfeld} 
lie just below the Fermi region.  This small discrepancy can be alleviated 
in many ways.  For example, the local dark matter density may be larger than the value 
assumed here, 0.43 GeV/cm$^3$.  
Moreover, taking into account substructure which enhances the positron and 
electron flux can lower the required cross-section by another  
factor of two \cite{Kuhlen:2009is,Cline:2010ag}.  
Uncertainties in the astrophysical backgrounds 
and in the cosmic-ray propagation can further lower the required cross-section and remove the 
discrepancy (see e.g.~\cite{Stawarz:2009ig,Katz:2009yd}).
In addition, the PAMELA/Fermi regions assume that $\phi$ annihilates exclusively to $\mu^+\mu^-$.  
In many models, however, $\phi$ will decay also at least partly to $e^+e^-$, which lowers the 
required cross-section further, while still providing an excellent fit to the data. 
This is the case, for example, in which dark matter is charged under a new GeV-scale Abelian 
force that kinetically mixes with hypercharge, see e.g.~\cite{Finkbeiner:2007kk, Arkani-Hamed2008, 
Pospelov:2008jd,Cholis:2008qq,Cholis2008,Finkbeiner:2010}.  

The next generation of ACTs such as AGIS~\footnote{www.agis-observatory.org} and 
CTA~\cite{Bringmann:2008kj} will improve the current 
prospects by roughly an order of magnitude, and will significantly constrain Sommerfeld 
enhanced models that explain the Fermi/PAMELA signals.  
The current and future prospects from gamma-ray observations of dwarf galaxies 
are complementary to those coming from current constraints of 
WMAP observations of the Cosmic Microwave Background and prospective constraints 
from Planck \cite{Padmanabhan:2005es,Galli:2009zc,Slatyer:2009yq,Galli:2010it}.

\section{Conclusions}\label{sec:concl}

The Milky Way satellite Segue 1 is the least luminous and most dark-matter dominated
galaxy known.  Using a new sample of member stars, we have derived a new 
value for the line-of-sight integral of the dark matter halo density-squared, $\Lann$.  
Its mean value is similar to the same quantity for Draco and larger than that 
for all other known dwarfs.
With this new value, we have derived robust upper limits on dark matter annihilation 
cross-sections from gamma-ray flux upper limits from 
Fermi and the Imaging Atmospheric Cherenkov Telescope (ACT) MAGIC, and  
have shown that these constraints may in fact be much stronger given the 
significant uncertainty remaining in the Segue 1 dark matter distribution. 
For a mean value of the Segue 1 density-squared integral, the constraints we 
present are more stringent than those obtained from any other Milky Way dwarf spheroidals
to date (see e.g.~\cite{Albert2008,Abdo:2010ex,Veritas:2010pj}).
 
We showed that Fermi and ACTs 
are complementary in the dark matter models they can probe, with 
Fermi having an advantage over ACTs for dark matter channels that produce softer photons as well as 
for lower dark matter masses $\lesssim \mathcal{O}$(100 GeV).  ACTs have an advantage over Fermi in 
constraining cross-sections for larger dark matter masses $\gtrsim \mathcal{O}$(100 GeV). 

The flux limits do not rule out a dark matter explanation for the 
PAMELA/Fermi/HESS $e^+/e^-$ signals, but probe the relevant regions.
With its low dark matter velocity dispersion, Segue 1 also serves as a useful 
laboratory for constraining Sommerfeld enhanced models that have been invoked 
to explain these signals. 
We find that current constraints from MAGIC slightly disfavor light 
mediators of mass $\mathcal{O}$(0.1 GeV) if we take the mean value 
for $\Lann$. 
However, due to the uncertainty in the dark matter density in Segue 1, 
such light force carriers cannot at this stage be ruled out.  

In addition, we show that under favorable circumstances neutrino observatories such as 
IceCube also have the potential to detect a dark matter 
signal from Segue 1, providing an important cross-check, for some channels, if a signal is 
detected in the gamma-rays.  

Since Segue 1 lies in the Northern Hemisphere, it is an excellent target for the VERITAS 
and MAGIC, and for IceCube.  
Observations by MAGIC (and future observations by VERITAS) 
are able to set interesting and robust constraints on dark matter annihilation, 
with rather modest observing times directed at Segue 1.  
Moreover, the ACT constraints presented here may be considerably improved if these instruments are able to lower 
their energy thresholds, observe over longer times, and perform an analysis specifically aimed at 
models that explain the $e^+/e^-$ signals.  

Though our analysis has focused on the galaxy Segue 1, prospects are promising for strengthening 
the limits we have presented. Given selection effects and biases in current surveys for 
MW satellites, it is probable that the population of nearby, dark-matter dominated satellites will
increase in the near future~\cite{Willman:2009dv}.  
Any signal from one dwarf can thus be corroborated by looking at others.  
Moreover, the significance of a signal or of a limit can be improved by stacking together the observations 
of several dwarf galaxies.  In addition, the next generation of ACTs such as 
AGIS and CTA will improve the current 
prospects by roughly an order of magnitude.  
These improvements will allow a definitive probe of many 
dark matter annihilation models. 

\acknowledgments

We thank E.~Bloom, D.~Cowen, K.~Hoffman, 
G.~Kane, M.~Kaplinghat, R.~Lu, S.~Murgia, E.~Resconi, J.~Shelton, P.~Schuster, 
N.~Toro, T.~Volansky, J.~Wacker for useful discussions or correspondence, 
and especially T.~Slatyer and N.~Weiner for useful discussions and comments on the draft.  
RE is supported by the US DOE under contract number DE-AC02-76SF00515.
NS is supported by the U.S. Department of Energy contract to SLAC no. DE-AC3-76SF00515.
LES acknowledges support for this work from NASA through Hubble Fellowship grant
HF-01225.01 awarded by the Space Telescope Science Institute, which is
operated by the Association of Universities for Research in Astronomy, Inc.,
for NASA, under contract NAS 5-26555.

\appendix

\section{Sommerfeld enhancement in the presence of an excited dark matter state}\label{app:inelastic}

In this brief Appendix, we summarize the formulas needed to calculate the Sommerfeld 
enhancement in the presence of an excited state \cite{Slatyer:2009vg}.  
The Sommerfeld effect can be calculated numerically, however \cite{Slatyer:2009vg} 
presented a computationally less intensive semi-analytical formula, which we review here. 
We refer the reader to \cite{Slatyer:2009vg} for all the details, and mostly use the same notation.   

We take the mass-splitting between the dark matter ground state and excited state to be 
$\delta$, and define
\bea
\epsilon_v = \f{v}{\alpha_D c}, \;\; \epsilon_\phi = \f{m_\phi}{\alpha_D m_\chi}, \;\;
\epsilon_\delta = \sqrt{\f{2 \delta}{m_\chi}}\, \f{1}{\alpha_D},
\eea
For $\epsilon_v, \epsilon_\phi, \epsilon_\delta \ll 1$ and for 
$\delta \lesssim$ max($\alpha_D m_\phi, m_\chi (v/c)^2$), the Sommerfeld enhanced 
annihilation of the ground-state is given by 
\begin{widetext}
\begin{equation}
S(v) = \f{2\pi}{\epsilon_v} \sinh\Big(\f{\pi \epsilon_v}{\mu}\Big) \times 
\begin{cases}
\f{1}{\cosh\left(\f{\pi \epsilon_v}{\mu}\right) - \cos\left(\f{\pi}{\mu} \sqrt{\epsilon_\delta^2-\epsilon_v^2} + 2\theta_-\right)}
& \epsilon_v < \epsilon_\delta, \\
& \\
\f{\cosh\Big(\f{\pi}{2\mu}\big(\epsilon_v+\sqrt{-\epsilon_\delta^2+\epsilon_v^2}\big)\Big) 
{\rm sech}\Big(\f{\pi}{2\mu}\big(\epsilon_v-\sqrt{-\epsilon_\delta^2+\epsilon_v^2}\big)\Big)}{
\cosh\Big(\f{\pi}{\mu}\big(\epsilon_v+\sqrt{-\epsilon_\delta^2+\epsilon_v^2}\big)\Big) - \cos(2\theta_-)} 
& \epsilon_v > \epsilon_\delta.
\end{cases}
\end{equation}
Here 
\bea
\mu & = & \epsilon_\phi \bigg(\f{1}{2} + \f{1}{2} \sqrt{1+\f{4}{\epsilon_\phi r_M}} \bigg) \\
\theta_- & = & \int_{z_M}^\infty \, \f{dz}{2\mu z} \,
\left( \sqrt{\bigg|\epsilon_v^2-\f{1}{2}\epsilon_\delta^2 + \sqrt{\Big(\f{1}{2}\epsilon^2_\delta\Big)^2 + 16 \mu^4 z}\bigg|} - 2 \mu z^{\f{1}{4}} \right) 
- 4 z_M^{\f{1}{4}} - \int_0^{r_M} \, dr\,  \sqrt{|-\lambda_-|} 
\eea
where $r_M$ and $z_M$ are defined by 
\bea
\f{e^{-\epsilon_\phi r_M}}{r_M} & = & \max\Big(\f{\epsilon_\delta^2}{2},\epsilon_\phi^2 \Big) \\
z_M & = & \max\Big(\f{\epsilon_\delta^4}{64 \mu^4}, \f{\epsilon_\phi^4}{16 \mu^4}\Big),
\eea
and $\lambda_-$ is given by
\bea
\lambda_-(r) = -\epsilon_v^2 + \f{\epsilon_\delta^2}{2} - 
\sqrt{\Big(\f{1}{2}\epsilon^2_\delta\Big)^2 + \f{e^{-2\epsilon_\phi r}}{r^2}}.
\eea
\end{widetext}

In the lower two figures in Figure \ref{fig:Sommerfeld}, we show the Sommerfeld enhanced 
annihilation cross-section for $\delta=1$ MeV.  Note that here we did not thermally average 
the cross-section, which is computationally very intensive, but rather took 150 km/s for 
the Milky-Way dark matter velocity and 4.5 km/s for that of Segue 1.  For the region of interest 
around $m_\chi \sim \mathcal{O}$(1 TeV), we checked that the resulting cross-section is 
$\sim  20\%$ \emph{lower} than the thermally averaged cross-section.  

\bibliography{SegueRefs} 

\begin{thebibliography}{97}
\expandafter\ifx\csname natexlab\endcsname\relax\def\natexlab#1{#1}\fi
\expandafter\ifx\csname bibnamefont\endcsname\relax
  \def\bibnamefont#1{#1}\fi
\expandafter\ifx\csname bibfnamefont\endcsname\relax
  \def\bibfnamefont#1{#1}\fi
\expandafter\ifx\csname citenamefont\endcsname\relax
  \def\citenamefont#1{#1}\fi
\expandafter\ifx\csname url\endcsname\relax
  \def\url#1{\texttt{#1}}\fi
\expandafter\ifx\csname urlprefix\endcsname\relax\def\urlprefix{URL }\fi
\providecommand{\bibinfo}[2]{#2}
\providecommand{\eprint}[2][]{\url{#2}}

\bibitem[{\citenamefont{Belokurov et~al.}(2007)}]{Belokurov:2006ph}
\bibinfo{author}{\bibfnamefont{V.}~\bibnamefont{Belokurov}}
  \bibnamefont{et~al.} (\bibinfo{collaboration}{SDSS}),
  \bibinfo{journal}{Astrophys. J.} \textbf{\bibinfo{volume}{654}},
  \bibinfo{pages}{897} (\bibinfo{year}{2007}), \eprint{astro-ph/0608448}.

\bibitem[{\citenamefont{{Geha} et~al.}(2009)\citenamefont{{Geha}, {Willman},
  {Simon}, {Strigari}, {Kirby}, {Law}, and {Strader}}}]{Geha:2008zr}
\bibinfo{author}{\bibfnamefont{M.}~\bibnamefont{{Geha}}},
  \bibinfo{author}{\bibfnamefont{B.}~\bibnamefont{{Willman}}},
  \bibinfo{author}{\bibfnamefont{J.~D.} \bibnamefont{{Simon}}},
  \bibinfo{author}{\bibfnamefont{L.~E.} \bibnamefont{{Strigari}}},
  \bibinfo{author}{\bibfnamefont{E.~N.} \bibnamefont{{Kirby}}},
  \bibinfo{author}{\bibfnamefont{D.~R.} \bibnamefont{{Law}}}, \bibnamefont{and}
  \bibinfo{author}{\bibfnamefont{J.}~\bibnamefont{{Strader}}},
  \bibinfo{journal}{\apj} \textbf{\bibinfo{volume}{692}}, \bibinfo{pages}{1464}
  (\bibinfo{year}{2009}), \eprint{0809.2781}.

\bibitem[{\citenamefont{Niederste-Ostholt
  et~al.}(2009)}]{NiedersteOstholt:2009na}
\bibinfo{author}{\bibfnamefont{M.}~\bibnamefont{Niederste-Ostholt}}
  \bibnamefont{et~al.} (\bibinfo{year}{2009}), \eprint{0906.3669}.

\bibitem[{\citenamefont{Simon et~al.}(2010)}]{Simon:2010ek}
\bibinfo{author}{\bibfnamefont{J.~D.} \bibnamefont{Simon}} \bibnamefont{et~al.}
  (\bibinfo{year}{2010}), \eprint{1007.4198}.

\bibitem[{\citenamefont{{M.~Geha}}()}]{Geha2009}
\bibinfo{author}{\bibnamefont{{M.~Geha}}}, \bibinfo{howpublished}{TeV Particle
  Astrophysics, 2009}.

\bibitem[{\citenamefont{Bergstrom et~al.}(2005)\citenamefont{Bergstrom,
  Bringmann, Eriksson, and Gustafsson}}]{Bergstrom:2004cy}
\bibinfo{author}{\bibfnamefont{L.}~\bibnamefont{Bergstrom}},
  \bibinfo{author}{\bibfnamefont{T.}~\bibnamefont{Bringmann}},
  \bibinfo{author}{\bibfnamefont{M.}~\bibnamefont{Eriksson}}, \bibnamefont{and}
  \bibinfo{author}{\bibfnamefont{M.}~\bibnamefont{Gustafsson}},
  \bibinfo{journal}{Phys. Rev. Lett.} \textbf{\bibinfo{volume}{94}},
  \bibinfo{pages}{131301} (\bibinfo{year}{2005}), \eprint{astro-ph/0410359}.

\bibitem[{\citenamefont{Beacom et~al.}(2005)\citenamefont{Beacom, Bell, and
  Bertone}}]{Beacom:2004pe}
\bibinfo{author}{\bibfnamefont{J.~F.} \bibnamefont{Beacom}},
  \bibinfo{author}{\bibfnamefont{N.~F.} \bibnamefont{Bell}}, \bibnamefont{and}
  \bibinfo{author}{\bibfnamefont{G.}~\bibnamefont{Bertone}},
  \bibinfo{journal}{Phys. Rev. Lett.} \textbf{\bibinfo{volume}{94}},
  \bibinfo{pages}{171301} (\bibinfo{year}{2005}), \eprint{astro-ph/0409403}.

\bibitem[{\citenamefont{{Birkedal} et~al.}(2005)}]{Birkedal2005}
\bibinfo{author}{\bibfnamefont{A.}~\bibnamefont{{Birkedal}}}
  \bibnamefont{et~al.} (\bibinfo{year}{2005}), \eprint{arXiv:hep-ph/0507194}.

\bibitem[{\citenamefont{Adriani et~al.}(2008{\natexlab{a}})}]{Adriani2008}
\bibinfo{author}{\bibfnamefont{O.}~\bibnamefont{Adriani}} \bibnamefont{et~al.}
  (\bibinfo{year}{2008}{\natexlab{a}}), \eprint{0810.4995}.

\bibitem[{\citenamefont{Abdo et~al.}(2009)}]{Abdo:2009zk}
\bibinfo{author}{\bibfnamefont{A.~A.} \bibnamefont{Abdo}} \bibnamefont{et~al.}
  (\bibinfo{collaboration}{The Fermi LAT}) (\bibinfo{year}{2009}),
  \eprint{0905.0025}.

\bibitem[{\citenamefont{Aharonian et~al.}(2008)}]{Collaboration:2008aaa}
\bibinfo{author}{\bibfnamefont{F.}~\bibnamefont{Aharonian}}
  \bibnamefont{et~al.} (\bibinfo{collaboration}{H.E.S.S.}),
  \bibinfo{journal}{Phys. Rev. Lett.} \textbf{\bibinfo{volume}{101}},
  \bibinfo{pages}{261104} (\bibinfo{year}{2008}), \eprint{0811.3894}.

\bibitem[{\citenamefont{Aharonian}(2009)}]{Aharonian:2009ah}
\bibinfo{author}{\bibfnamefont{H.~E. S. S. C.~F.} \bibnamefont{Aharonian}}
  (\bibinfo{year}{2009}), \eprint{0905.0105}.

\bibitem[{\citenamefont{Adriani et~al.}(2008{\natexlab{b}})}]{Adriani:2008zq}
\bibinfo{author}{\bibfnamefont{O.}~\bibnamefont{Adriani}} \bibnamefont{et~al.}
  (\bibinfo{year}{2008}{\natexlab{b}}), \eprint{0810.4994}.

\bibitem[{\citenamefont{Cirelli et~al.}(2008)}]{Cirelli:2008pk}
\bibinfo{author}{\bibfnamefont{M.}~\bibnamefont{Cirelli}} \bibnamefont{et~al.}
  (\bibinfo{year}{2008}), \eprint{0809.2409}.

\bibitem[{\citenamefont{{Arkani-Hamed} et~al.}(2008)}]{Arkani-Hamed2008}
\bibinfo{author}{\bibfnamefont{N.}~\bibnamefont{{Arkani-Hamed}}}
  \bibnamefont{et~al.} (\bibinfo{year}{2008}), \eprint{0810.0713}.

\bibitem[{\citenamefont{Pospelov and Ritz}(2009)}]{Pospelov:2008jd}
\bibinfo{author}{\bibfnamefont{M.}~\bibnamefont{Pospelov}} \bibnamefont{and}
  \bibinfo{author}{\bibfnamefont{A.}~\bibnamefont{Ritz}},
  \bibinfo{journal}{Phys. Lett.} \textbf{\bibinfo{volume}{B671}},
  \bibinfo{pages}{391} (\bibinfo{year}{2009}), \eprint{0810.1502}.

\bibitem[{\citenamefont{Cholis et~al.}(2008{\natexlab{a}})\citenamefont{Cholis,
  Finkbeiner, Goodenough, and Weiner}}]{Cholis:2008qq}
\bibinfo{author}{\bibfnamefont{I.}~\bibnamefont{Cholis}},
  \bibinfo{author}{\bibfnamefont{D.~P.} \bibnamefont{Finkbeiner}},
  \bibinfo{author}{\bibfnamefont{L.}~\bibnamefont{Goodenough}},
  \bibnamefont{and} \bibinfo{author}{\bibfnamefont{N.}~\bibnamefont{Weiner}}
  (\bibinfo{year}{2008}{\natexlab{a}}), \eprint{0810.5344}.

\bibitem[{\citenamefont{Cholis et~al.}(2008{\natexlab{b}})\citenamefont{Cholis,
  Dobler, Finkbeiner, Goodenough, and Weiner}}]{Cholis2008}
\bibinfo{author}{\bibfnamefont{I.}~\bibnamefont{Cholis}},
  \bibinfo{author}{\bibfnamefont{G.}~\bibnamefont{Dobler}},
  \bibinfo{author}{\bibfnamefont{D.~P.} \bibnamefont{Finkbeiner}},
  \bibinfo{author}{\bibfnamefont{L.}~\bibnamefont{Goodenough}},
  \bibnamefont{and} \bibinfo{author}{\bibfnamefont{N.}~\bibnamefont{Weiner}}
  (\bibinfo{year}{2008}{\natexlab{b}}), \eprint{0811.3641}.

\bibitem[{\citenamefont{Bergstrom et~al.}(2009)\citenamefont{Bergstrom, Edsjo,
  and Zaharijas}}]{Bergstrom:2009fa}
\bibinfo{author}{\bibfnamefont{L.}~\bibnamefont{Bergstrom}},
  \bibinfo{author}{\bibfnamefont{J.}~\bibnamefont{Edsjo}}, \bibnamefont{and}
  \bibinfo{author}{\bibfnamefont{G.}~\bibnamefont{Zaharijas}}
  (\bibinfo{year}{2009}), \eprint{0905.0333}.

\bibitem[{\citenamefont{Meade et~al.}(2009)\citenamefont{Meade, Papucci,
  Strumia, and Volansky}}]{Meade:2009iu}
\bibinfo{author}{\bibfnamefont{P.}~\bibnamefont{Meade}},
  \bibinfo{author}{\bibfnamefont{M.}~\bibnamefont{Papucci}},
  \bibinfo{author}{\bibfnamefont{A.}~\bibnamefont{Strumia}}, \bibnamefont{and}
  \bibinfo{author}{\bibfnamefont{T.}~\bibnamefont{Volansky}}
  (\bibinfo{year}{2009}), \eprint{0905.0480}.

\bibitem[{\citenamefont{Martinez et~al.}(2009)\citenamefont{Martinez, Bullock,
  Kaplinghat, Strigari, and Trotta}}]{Martinez:2009jh}
\bibinfo{author}{\bibfnamefont{G.~D.} \bibnamefont{Martinez}},
  \bibinfo{author}{\bibfnamefont{J.~S.} \bibnamefont{Bullock}},
  \bibinfo{author}{\bibfnamefont{M.}~\bibnamefont{Kaplinghat}},
  \bibinfo{author}{\bibfnamefont{L.~E.} \bibnamefont{Strigari}},
  \bibnamefont{and} \bibinfo{author}{\bibfnamefont{R.}~\bibnamefont{Trotta}},
  \bibinfo{journal}{JCAP} \textbf{\bibinfo{volume}{0906}}, \bibinfo{pages}{014}
  (\bibinfo{year}{2009}), \eprint{0902.4715}.

\bibitem[{\citenamefont{Essig et~al.}(2009)\citenamefont{Essig, Sehgal, and
  Strigari}}]{Essig:2009jx}
\bibinfo{author}{\bibfnamefont{R.}~\bibnamefont{Essig}},
  \bibinfo{author}{\bibfnamefont{N.}~\bibnamefont{Sehgal}}, \bibnamefont{and}
  \bibinfo{author}{\bibfnamefont{L.~E.} \bibnamefont{Strigari}},
  \bibinfo{journal}{Phys. Rev.} \textbf{\bibinfo{volume}{D80}},
  \bibinfo{pages}{023506} (\bibinfo{year}{2009}), \eprint{0902.4750}.

\bibitem[{\citenamefont{Scott et~al.}(2010)}]{Scott:2009jn}
\bibinfo{author}{\bibfnamefont{P.}~\bibnamefont{Scott}} \bibnamefont{et~al.},
  \bibinfo{journal}{JCAP} \textbf{\bibinfo{volume}{1001}}, \bibinfo{pages}{031}
  (\bibinfo{year}{2010}), \eprint{0909.3300}.

\bibitem[{\citenamefont{Perelstein and Shakya}(2010)}]{Perelstein:2010at}
\bibinfo{author}{\bibfnamefont{M.}~\bibnamefont{Perelstein}} \bibnamefont{and}
  \bibinfo{author}{\bibfnamefont{B.}~\bibnamefont{Shakya}}
  (\bibinfo{year}{2010}), \eprint{1007.0018}.

\bibitem[{\citenamefont{Minor et~al.}(2010)\citenamefont{Minor, Martinez,
  Bullock, Kaplinghat, and Trainor}}]{Minor:2010vp}
\bibinfo{author}{\bibfnamefont{Q.~E.} \bibnamefont{Minor}},
  \bibinfo{author}{\bibfnamefont{G.}~\bibnamefont{Martinez}},
  \bibinfo{author}{\bibfnamefont{J.}~\bibnamefont{Bullock}},
  \bibinfo{author}{\bibfnamefont{M.}~\bibnamefont{Kaplinghat}},
  \bibnamefont{and} \bibinfo{author}{\bibfnamefont{R.}~\bibnamefont{Trainor}}
  (\bibinfo{year}{2010}), \eprint{1001.1160}.

\bibitem[{\citenamefont{{Strigari} et~al.}(2008)}]{Strigari:2007at}
\bibinfo{author}{\bibfnamefont{L.~E.} \bibnamefont{{Strigari}}}
  \bibnamefont{et~al.}, \bibinfo{journal}{\apj} \textbf{\bibinfo{volume}{678}},
  \bibinfo{pages}{614} (\bibinfo{year}{2008}), \eprint{0709.1510}.

\bibitem[{\citenamefont{{Navarro} et~al.}(2008)}]{Navarro:2008kc}
\bibinfo{author}{\bibfnamefont{J.~F.} \bibnamefont{{Navarro}}}
  \bibnamefont{et~al.} (\bibinfo{year}{2008}), \eprint{0810.1522}.

\bibitem[{\citenamefont{Regis and Ullio}(2009)}]{Regis:2009qt}
\bibinfo{author}{\bibfnamefont{M.}~\bibnamefont{Regis}} \bibnamefont{and}
  \bibinfo{author}{\bibfnamefont{P.}~\bibnamefont{Ullio}}
  (\bibinfo{year}{2009}), \eprint{0907.5093}.

\bibitem[{\citenamefont{Colin et~al.}(2000)\citenamefont{Colin, Avila-Reese,
  and Valenzuela}}]{Colin2000}
\bibinfo{author}{\bibfnamefont{P.}~\bibnamefont{Colin}},
  \bibinfo{author}{\bibfnamefont{V.}~\bibnamefont{Avila-Reese}},
  \bibnamefont{and}
  \bibinfo{author}{\bibfnamefont{O.}~\bibnamefont{Valenzuela}},
  \bibinfo{journal}{Astrophys. J.} \textbf{\bibinfo{volume}{542}},
  \bibinfo{pages}{622} (\bibinfo{year}{2000}), \eprint{astro-ph/0004115}.

\bibitem[{\citenamefont{{Widrow} and {Dubinski}}(2005)}]{Widrow2005}
\bibinfo{author}{\bibfnamefont{L.~M.} \bibnamefont{{Widrow}}} \bibnamefont{and}
  \bibinfo{author}{\bibfnamefont{J.}~\bibnamefont{{Dubinski}}},
  \bibinfo{journal}{\apj} \textbf{\bibinfo{volume}{631}}, \bibinfo{pages}{838}
  (\bibinfo{year}{2005}), \eprint{arXiv:astro-ph/0506177}.

\bibitem[{\citenamefont{Abdo et~al.}(2010{\natexlab{a}})}]{Abdo:2010ex}
\bibinfo{author}{\bibfnamefont{A.~A.} \bibnamefont{Abdo}} \bibnamefont{et~al.},
  \bibinfo{journal}{Astrophys. J.} \textbf{\bibinfo{volume}{712}},
  \bibinfo{pages}{147} (\bibinfo{year}{2010}{\natexlab{a}}),
  \eprint{1001.4531}.

\bibitem[{\citenamefont{Willman et~al.}(2010)}]{Willman:2010gy}
\bibinfo{author}{\bibfnamefont{B.}~\bibnamefont{Willman}} \bibnamefont{et~al.}
  (\bibinfo{year}{2010}), \eprint{1007.3499}.

\bibitem[{\citenamefont{Ibata et~al.}(1995)\citenamefont{Ibata, Gilmore, and
  Irwin}}]{Ibata:1995fz}
\bibinfo{author}{\bibfnamefont{R.~A.} \bibnamefont{Ibata}},
  \bibinfo{author}{\bibfnamefont{G.~F.} \bibnamefont{Gilmore}},
  \bibnamefont{and} \bibinfo{author}{\bibfnamefont{M.~J.} \bibnamefont{Irwin}},
  \bibinfo{journal}{Mon. Not. Roy. Astron. Soc.}
  \textbf{\bibinfo{volume}{277}}, \bibinfo{pages}{781} (\bibinfo{year}{1995}),
  \eprint{astro-ph/9506071}.

\bibitem[{\citenamefont{Mack et~al.}(2008)}]{Mack:2008wu}
\bibinfo{author}{\bibfnamefont{G.~D.} \bibnamefont{Mack}} \bibnamefont{et~al.},
  \bibinfo{journal}{Phys. Rev.} \textbf{\bibinfo{volume}{D78}},
  \bibinfo{pages}{063542} (\bibinfo{year}{2008}), \eprint{0803.0157}.

\bibitem[{\citenamefont{{Bell} and {Jacques}}(2008)}]{Bell2008}
\bibinfo{author}{\bibfnamefont{N.~F.} \bibnamefont{{Bell}}} \bibnamefont{and}
  \bibinfo{author}{\bibfnamefont{T.~D.} \bibnamefont{{Jacques}}}
  (\bibinfo{year}{2008}), \eprint{0811.0821}.

\bibitem[{\citenamefont{Fortin et~al.}(2009)\citenamefont{Fortin, Shelton,
  Thomas, and Zhao}}]{Fortin:2009rq}
\bibinfo{author}{\bibfnamefont{J.-F.} \bibnamefont{Fortin}},
  \bibinfo{author}{\bibfnamefont{J.}~\bibnamefont{Shelton}},
  \bibinfo{author}{\bibfnamefont{S.}~\bibnamefont{Thomas}}, \bibnamefont{and}
  \bibinfo{author}{\bibfnamefont{Y.}~\bibnamefont{Zhao}}
  (\bibinfo{year}{2009}), \eprint{0908.2258}.

\bibitem[{\citenamefont{Fornengo et~al.}(2004)\citenamefont{Fornengo, Pieri,
  and Scopel}}]{Fornengo:2004kj}
\bibinfo{author}{\bibfnamefont{N.}~\bibnamefont{Fornengo}},
  \bibinfo{author}{\bibfnamefont{L.}~\bibnamefont{Pieri}}, \bibnamefont{and}
  \bibinfo{author}{\bibfnamefont{S.}~\bibnamefont{Scopel}},
  \bibinfo{journal}{Phys. Rev.} \textbf{\bibinfo{volume}{D70}},
  \bibinfo{pages}{103529} (\bibinfo{year}{2004}), \eprint{hep-ph/0407342}.

\bibitem[{\citenamefont{Kuno and Okada}(2001)}]{Kuno:1999jp}
\bibinfo{author}{\bibfnamefont{Y.}~\bibnamefont{Kuno}} \bibnamefont{and}
  \bibinfo{author}{\bibfnamefont{Y.}~\bibnamefont{Okada}},
  \bibinfo{journal}{Rev. Mod. Phys.} \textbf{\bibinfo{volume}{73}},
  \bibinfo{pages}{151} (\bibinfo{year}{2001}), \eprint{hep-ph/9909265}.

\bibitem[{\citenamefont{{Mardon} et~al.}(2009)}]{Mardon2009}
\bibinfo{author}{\bibfnamefont{J.}~\bibnamefont{{Mardon}}} \bibnamefont{et~al.}
  (\bibinfo{year}{2009}), \eprint{0901.2926}.

\bibitem[{\citenamefont{Kane et~al.}(2009)\citenamefont{Kane, Lu, and
  Watson}}]{Kane:2009if}
\bibinfo{author}{\bibfnamefont{G.}~\bibnamefont{Kane}},
  \bibinfo{author}{\bibfnamefont{R.}~\bibnamefont{Lu}}, \bibnamefont{and}
  \bibinfo{author}{\bibfnamefont{S.}~\bibnamefont{Watson}}
  (\bibinfo{year}{2009}), \eprint{0906.4765}.

\bibitem[{\citenamefont{Gondolo et~al.}(2004)}]{Gondolo:2004sc}
\bibinfo{author}{\bibfnamefont{P.}~\bibnamefont{Gondolo}} \bibnamefont{et~al.},
  \bibinfo{journal}{JCAP} \textbf{\bibinfo{volume}{0407}}, \bibinfo{pages}{008}
  (\bibinfo{year}{2004}), \eprint{astro-ph/0406204}.

\bibitem[{\citenamefont{{http://www.physto.se/$\sim$edsjo/darksusy}}()}]{darks%
usy}
\bibinfo{author}{\bibnamefont{{http://www.physto.se/$\sim$edsjo/darksusy}}}.

\bibitem[{\citenamefont{Barger et~al.}(2007)\citenamefont{Barger, Keung,
  Shaughnessy, and Tregre}}]{Barger:2007xf}
\bibinfo{author}{\bibfnamefont{V.}~\bibnamefont{Barger}},
  \bibinfo{author}{\bibfnamefont{W.-Y.} \bibnamefont{Keung}},
  \bibinfo{author}{\bibfnamefont{G.}~\bibnamefont{Shaughnessy}},
  \bibnamefont{and} \bibinfo{author}{\bibfnamefont{A.}~\bibnamefont{Tregre}},
  \bibinfo{journal}{Phys. Rev.} \textbf{\bibinfo{volume}{D76}},
  \bibinfo{pages}{095008} (\bibinfo{year}{2007}), \eprint{0708.1325}.

\bibitem[{\citenamefont{Sandick et~al.}(2010)\citenamefont{Sandick, Spolyar,
  Buckley, Freese, and Hooper}}]{Sandick:2009bi}
\bibinfo{author}{\bibfnamefont{P.}~\bibnamefont{Sandick}},
  \bibinfo{author}{\bibfnamefont{D.}~\bibnamefont{Spolyar}},
  \bibinfo{author}{\bibfnamefont{M.~R.} \bibnamefont{Buckley}},
  \bibinfo{author}{\bibfnamefont{K.}~\bibnamefont{Freese}}, \bibnamefont{and}
  \bibinfo{author}{\bibfnamefont{D.}~\bibnamefont{Hooper}},
  \bibinfo{journal}{Phys. Rev.} \textbf{\bibinfo{volume}{D81}},
  \bibinfo{pages}{083506} (\bibinfo{year}{2010}), \eprint{0912.0513}.

\bibitem[{\citenamefont{Amsler et~al.}(2008)}]{Amsler:2008zzb}
\bibinfo{author}{\bibfnamefont{C.}~\bibnamefont{Amsler}} \bibnamefont{et~al.}
  (\bibinfo{collaboration}{Particle Data Group}), \bibinfo{journal}{Phys.
  Lett.} \textbf{\bibinfo{volume}{B667}}, \bibinfo{pages}{1}
  (\bibinfo{year}{2008}).

\bibitem[{\citenamefont{Resconi
  (IceCube~Collaboration)}(2009)}]{Resconi:2008fe}
\bibinfo{author}{\bibfnamefont{E.}~\bibnamefont{Resconi
  (IceCube~Collaboration)}}, \bibinfo{journal}{Nucl. Instrum. Meth.}
  \textbf{\bibinfo{volume}{A602}}, \bibinfo{pages}{7} (\bibinfo{year}{2009}),
  \eprint{0807.3891}.

\bibitem[{\citenamefont{Salucci et~al.}(2010)\citenamefont{Salucci, Nesti,
  Gentile, and Martins}}]{Salucci:2010qr}
\bibinfo{author}{\bibfnamefont{P.}~\bibnamefont{Salucci}},
  \bibinfo{author}{\bibfnamefont{F.}~\bibnamefont{Nesti}},
  \bibinfo{author}{\bibfnamefont{G.}~\bibnamefont{Gentile}}, \bibnamefont{and}
  \bibinfo{author}{\bibfnamefont{C.~F.} \bibnamefont{Martins}}
  (\bibinfo{year}{2010}), \eprint{1003.3101}.

\bibitem[{\citenamefont{Barger and Phillips}(1987)}]{Barger:1987}
\bibinfo{author}{\bibfnamefont{V.}~\bibnamefont{Barger}} \bibnamefont{and}
  \bibinfo{author}{\bibfnamefont{R.}~\bibnamefont{Phillips}},
  \emph{\bibinfo{title}{Collider Physics}} (\bibinfo{publisher}{Addison-Wesley
  (Redwood City, USA)}, \bibinfo{year}{1987}).

\bibitem[{\citenamefont{Winter}(2000)}]{Winter:2000}
\bibinfo{author}{\bibfnamefont{K.}~\bibnamefont{Winter}},
  \emph{\bibinfo{title}{Neutrino Physics}} (\bibinfo{publisher}{Cambridge
  University Press}, \bibinfo{year}{2000}).

\bibitem[{\citenamefont{{Ahrens} et~al.}(2004)\citenamefont{{Ahrens},
  {Bahcall}, {Bai}, {Bay}, {Becka}, {Becker}, {Berley}, {Bernardini},
  {Bertrand}, {Besson} et~al.}}]{Ahrens2004}
\bibinfo{author}{\bibfnamefont{J.}~\bibnamefont{{Ahrens}}},
  \bibinfo{author}{\bibfnamefont{J.~N.} \bibnamefont{{Bahcall}}},
  \bibinfo{author}{\bibfnamefont{X.}~\bibnamefont{{Bai}}},
  \bibinfo{author}{\bibfnamefont{R.~C.} \bibnamefont{{Bay}}},
  \bibinfo{author}{\bibfnamefont{T.}~\bibnamefont{{Becka}}},
  \bibinfo{author}{\bibfnamefont{K.-H.} \bibnamefont{{Becker}}},
  \bibinfo{author}{\bibfnamefont{D.}~\bibnamefont{{Berley}}},
  \bibinfo{author}{\bibfnamefont{E.}~\bibnamefont{{Bernardini}}},
  \bibinfo{author}{\bibfnamefont{D.}~\bibnamefont{{Bertrand}}},
  \bibinfo{author}{\bibfnamefont{D.~Z.} \bibnamefont{{Besson}}},
  \bibnamefont{et~al.}, \bibinfo{journal}{Astroparticle Physics}
  \textbf{\bibinfo{volume}{20}}, \bibinfo{pages}{507} (\bibinfo{year}{2004}),
  \eprint{arXiv:astro-ph/0305196}.

\bibitem[{\citenamefont{Gonzalez-Garcia
  et~al.}(2005)\citenamefont{Gonzalez-Garcia, Halzen, and
  Maltoni}}]{GonzalezGarcia:2005xw}
\bibinfo{author}{\bibfnamefont{M.~C.} \bibnamefont{Gonzalez-Garcia}},
  \bibinfo{author}{\bibfnamefont{F.}~\bibnamefont{Halzen}}, \bibnamefont{and}
  \bibinfo{author}{\bibfnamefont{M.}~\bibnamefont{Maltoni}},
  \bibinfo{journal}{Phys. Rev.} \textbf{\bibinfo{volume}{D71}},
  \bibinfo{pages}{093010} (\bibinfo{year}{2005}), \eprint{hep-ph/0502223}.

\bibitem[{\citenamefont{Honda et~al.}(2007)\citenamefont{Honda, Kajita,
  Kasahara, Midorikawa, and Sanuki}}]{Honda:2006qj}
\bibinfo{author}{\bibfnamefont{M.}~\bibnamefont{Honda}},
  \bibinfo{author}{\bibfnamefont{T.}~\bibnamefont{Kajita}},
  \bibinfo{author}{\bibfnamefont{K.}~\bibnamefont{Kasahara}},
  \bibinfo{author}{\bibfnamefont{S.}~\bibnamefont{Midorikawa}},
  \bibnamefont{and} \bibinfo{author}{\bibfnamefont{T.}~\bibnamefont{Sanuki}},
  \bibinfo{journal}{Phys. Rev.} \textbf{\bibinfo{volume}{D75}},
  \bibinfo{pages}{043006} (\bibinfo{year}{2007}), \eprint{astro-ph/0611418}.

\bibitem[{\citenamefont{Ahrens et~al.}(2004)}]{Ahrens:2003ix}
\bibinfo{author}{\bibfnamefont{J.}~\bibnamefont{Ahrens}} \bibnamefont{et~al.}
  (\bibinfo{collaboration}{IceCube}), \bibinfo{journal}{Astropart. Phys.}
  \textbf{\bibinfo{volume}{20}}, \bibinfo{pages}{507} (\bibinfo{year}{2004}),
  \eprint{astro-ph/0305196}.

\bibitem[{\citenamefont{{P.~Wang (Fermi-LAT)}}()}]{Wang}
\bibinfo{author}{\bibnamefont{{P.~Wang (Fermi-LAT)}}},
  \bibinfo{howpublished}{Cosmology in Northern California Meeting, 2009}.

\bibitem[{\citenamefont{{C.~Farnier (Fermi-LAT)}}()}]{Farnier}
\bibinfo{author}{\bibnamefont{{C.~Farnier (Fermi-LAT)}}},
  \bibinfo{howpublished}{Roma International Conference on Astro-Particle
  Physics, 2009}.

\bibitem[{\citenamefont{{S.~Murgia (Fermi-LAT)}}()}]{Murgia}
\bibinfo{author}{\bibnamefont{{S.~Murgia (Fermi-LAT)}}},
  \bibinfo{howpublished}{TeV Particle Astrophysics, 2009}.

\bibitem[{\citenamefont{{T.~Jeltema (Fermi-LAT)}}()}]{Jeltema}
\bibinfo{author}{\bibnamefont{{T.~Jeltema (Fermi-LAT)}}},
  \bibinfo{howpublished}{TeV Particle Astrophysics, 2009}.

\bibitem[{\citenamefont{{D.N.~Casta\~no}}()}]{magic}
\bibinfo{author}{\bibnamefont{{D.N.~Casta\~no}}}, \bibinfo{howpublished}{Dark
  Matter 2010, UCLA, http://www.physics.ucla.edu/hep/dm10/talks/ \\
  nietocastano.pdf}.

\bibitem[{\citenamefont{Abdo et~al.}(2010{\natexlab{b}})}]{Abdo:2010nc}
\bibinfo{author}{\bibfnamefont{A.~A.} \bibnamefont{Abdo}} \bibnamefont{et~al.},
  \bibinfo{journal}{Phys. Rev. Lett.} \textbf{\bibinfo{volume}{104}},
  \bibinfo{pages}{091302} (\bibinfo{year}{2010}{\natexlab{b}}),
  \eprint{1001.4836}.

\bibitem[{\citenamefont{Spolyar et~al.}(2009)\citenamefont{Spolyar, Buckley,
  Freese, Hooper, and Murayama}}]{Spolyar:2009kx}
\bibinfo{author}{\bibfnamefont{D.}~\bibnamefont{Spolyar}},
  \bibinfo{author}{\bibfnamefont{M.~R.} \bibnamefont{Buckley}},
  \bibinfo{author}{\bibfnamefont{K.}~\bibnamefont{Freese}},
  \bibinfo{author}{\bibfnamefont{D.}~\bibnamefont{Hooper}}, \bibnamefont{and}
  \bibinfo{author}{\bibfnamefont{H.}~\bibnamefont{Murayama}}
  (\bibinfo{year}{2009}), \eprint{0905.4764}.

\bibitem[{\citenamefont{Kamionkowski and Turner}(1990)}]{Kamionkowski:1990}
\bibinfo{author}{\bibfnamefont{M.}~\bibnamefont{Kamionkowski}}
  \bibnamefont{and} \bibinfo{author}{\bibfnamefont{M.~S.}
  \bibnamefont{Turner}}, \bibinfo{journal}{Phys. Rev. D}
  \textbf{\bibinfo{volume}{42}}, \bibinfo{pages}{3310} (\bibinfo{year}{1990}).

\bibitem[{\citenamefont{Moroi and Randall}(2000)}]{Moroi:1999zb}
\bibinfo{author}{\bibfnamefont{T.}~\bibnamefont{Moroi}} \bibnamefont{and}
  \bibinfo{author}{\bibfnamefont{L.}~\bibnamefont{Randall}},
  \bibinfo{journal}{Nucl. Phys.} \textbf{\bibinfo{volume}{B570}},
  \bibinfo{pages}{455} (\bibinfo{year}{2000}), \eprint{hep-ph/9906527}.

\bibitem[{\citenamefont{Giudice et~al.}(2001)\citenamefont{Giudice, Kolb, and
  Riotto}}]{Giudice:2000ex}
\bibinfo{author}{\bibfnamefont{G.~F.} \bibnamefont{Giudice}},
  \bibinfo{author}{\bibfnamefont{E.~W.} \bibnamefont{Kolb}}, \bibnamefont{and}
  \bibinfo{author}{\bibfnamefont{A.}~\bibnamefont{Riotto}},
  \bibinfo{journal}{Phys. Rev.} \textbf{\bibinfo{volume}{D64}},
  \bibinfo{pages}{023508} (\bibinfo{year}{2001}), \eprint{hep-ph/0005123}.

\bibitem[{\citenamefont{Ibe et~al.}(2009)\citenamefont{Ibe, Murayama, and
  Yanagida}}]{Ibe:2008ye}
\bibinfo{author}{\bibfnamefont{M.}~\bibnamefont{Ibe}},
  \bibinfo{author}{\bibfnamefont{H.}~\bibnamefont{Murayama}}, \bibnamefont{and}
  \bibinfo{author}{\bibfnamefont{T.~T.} \bibnamefont{Yanagida}},
  \bibinfo{journal}{Phys. Rev.} \textbf{\bibinfo{volume}{D79}},
  \bibinfo{pages}{095009} (\bibinfo{year}{2009}), \eprint{0812.0072}.

\bibitem[{\citenamefont{Sommerfeld}(1931)}]{Sommerfeld}
\bibinfo{author}{\bibfnamefont{A.}~\bibnamefont{Sommerfeld}},
  \bibinfo{journal}{Annalen der Physik 403, 257}  (\bibinfo{year}{1931}).

\bibitem[{\citenamefont{Hisano et~al.}(2004)\citenamefont{Hisano, Matsumoto,
  and Nojiri}}]{Hisano:2003ec}
\bibinfo{author}{\bibfnamefont{J.}~\bibnamefont{Hisano}},
  \bibinfo{author}{\bibfnamefont{S.}~\bibnamefont{Matsumoto}},
  \bibnamefont{and} \bibinfo{author}{\bibfnamefont{M.~M.}
  \bibnamefont{Nojiri}}, \bibinfo{journal}{Phys. Rev. Lett.}
  \textbf{\bibinfo{volume}{92}}, \bibinfo{pages}{031303}
  (\bibinfo{year}{2004}), \eprint{hep-ph/0307216}.

\bibitem[{\citenamefont{March-Russell et~al.}(2008)\citenamefont{March-Russell,
  West, Cumberbatch, and Hooper}}]{MarchRussell:2008yu}
\bibinfo{author}{\bibfnamefont{J.}~\bibnamefont{March-Russell}},
  \bibinfo{author}{\bibfnamefont{S.~M.} \bibnamefont{West}},
  \bibinfo{author}{\bibfnamefont{D.}~\bibnamefont{Cumberbatch}},
  \bibnamefont{and} \bibinfo{author}{\bibfnamefont{D.}~\bibnamefont{Hooper}},
  \bibinfo{journal}{JHEP} \textbf{\bibinfo{volume}{07}}, \bibinfo{pages}{058}
  (\bibinfo{year}{2008}), \eprint{0801.3440}.

\bibitem[{\citenamefont{March-Russell and West}(2009)}]{MarchRussell:2008tu}
\bibinfo{author}{\bibfnamefont{J.~D.} \bibnamefont{March-Russell}}
  \bibnamefont{and} \bibinfo{author}{\bibfnamefont{S.~M.} \bibnamefont{West}},
  \bibinfo{journal}{Phys. Lett.} \textbf{\bibinfo{volume}{B676}},
  \bibinfo{pages}{133} (\bibinfo{year}{2009}), \eprint{0812.0559}.

\bibitem[{\citenamefont{Robertson and Zentner}(2009)}]{Robertson:2009bh}
\bibinfo{author}{\bibfnamefont{B.}~\bibnamefont{Robertson}} \bibnamefont{and}
  \bibinfo{author}{\bibfnamefont{A.}~\bibnamefont{Zentner}}
  (\bibinfo{year}{2009}), \eprint{0902.0362}.

\bibitem[{\citenamefont{Kuhlen}(2009)}]{Kuhlen:2009jv}
\bibinfo{author}{\bibfnamefont{M.}~\bibnamefont{Kuhlen}}
  (\bibinfo{year}{2009}), \eprint{0906.1822}.

\bibitem[{\citenamefont{Kuhlen et~al.}(2009)\citenamefont{Kuhlen, Madau, and
  Silk}}]{Kuhlen:2009kx}
\bibinfo{author}{\bibfnamefont{M.}~\bibnamefont{Kuhlen}},
  \bibinfo{author}{\bibfnamefont{P.}~\bibnamefont{Madau}}, \bibnamefont{and}
  \bibinfo{author}{\bibfnamefont{J.}~\bibnamefont{Silk}}
  (\bibinfo{year}{2009}), \eprint{0907.0005}.

\bibitem[{\citenamefont{Bovy}(2009)}]{Bovy:2009zs}
\bibinfo{author}{\bibfnamefont{J.}~\bibnamefont{Bovy}}, \bibinfo{journal}{Phys.
  Rev.} \textbf{\bibinfo{volume}{D79}}, \bibinfo{pages}{083539}
  (\bibinfo{year}{2009}), \eprint{0903.0413}.

\bibitem[{\citenamefont{Slatyer}(2010)}]{Slatyer:2009vg}
\bibinfo{author}{\bibfnamefont{T.~R.} \bibnamefont{Slatyer}},
  \bibinfo{journal}{JCAP} \textbf{\bibinfo{volume}{1002}}, \bibinfo{pages}{028}
  (\bibinfo{year}{2010}), \eprint{0910.5713}.

\bibitem[{\citenamefont{Tucker-Smith and Weiner}(2001)}]{TuckerSmith:2001hy}
\bibinfo{author}{\bibfnamefont{D.}~\bibnamefont{Tucker-Smith}}
  \bibnamefont{and} \bibinfo{author}{\bibfnamefont{N.}~\bibnamefont{Weiner}},
  \bibinfo{journal}{Phys. Rev.} \textbf{\bibinfo{volume}{D64}},
  \bibinfo{pages}{043502} (\bibinfo{year}{2001}), \eprint{hep-ph/0101138}.

\bibitem[{\citenamefont{Finkbeiner and Weiner}(2007)}]{Finkbeiner:2007kk}
\bibinfo{author}{\bibfnamefont{D.~P.} \bibnamefont{Finkbeiner}}
  \bibnamefont{and} \bibinfo{author}{\bibfnamefont{N.}~\bibnamefont{Weiner}},
  \bibinfo{journal}{Phys. Rev.} \textbf{\bibinfo{volume}{D76}},
  \bibinfo{pages}{083519} (\bibinfo{year}{2007}), \eprint{astro-ph/0702587}.

\bibitem[{\citenamefont{Feng et~al.}(2010{\natexlab{a}})\citenamefont{Feng,
  Kaplinghat, and Yu}}]{Feng:2010zp}
\bibinfo{author}{\bibfnamefont{J.~L.} \bibnamefont{Feng}},
  \bibinfo{author}{\bibfnamefont{M.}~\bibnamefont{Kaplinghat}},
  \bibnamefont{and} \bibinfo{author}{\bibfnamefont{H.-B.} \bibnamefont{Yu}}
  (\bibinfo{year}{2010}{\natexlab{a}}), \eprint{1005.4678}.

\bibitem[{\citenamefont{Finkbeiner et~al.}(2010)\citenamefont{Finkbeiner,
  Goodenough, Slatyer, and Weiner}}]{Finkbeiner:2010}
\bibinfo{author}{\bibfnamefont{D.~P.} \bibnamefont{Finkbeiner}},
  \bibinfo{author}{\bibfnamefont{L.}~\bibnamefont{Goodenough}},
  \bibinfo{author}{\bibfnamefont{T.}~\bibnamefont{Slatyer}}, \bibnamefont{and}
  \bibinfo{author}{\bibfnamefont{N.}~\bibnamefont{Weiner}},
  \bibinfo{journal}{to appear}  (\bibinfo{year}{2010}).

\bibitem[{\citenamefont{Chen et~al.}(2007)\citenamefont{Chen, Lu, and
  Sun}}]{Chen2007}
\bibinfo{author}{\bibfnamefont{C.-Y.} \bibnamefont{Chen}},
  \bibinfo{author}{\bibfnamefont{F.-L.} \bibnamefont{Lu}}, \bibnamefont{and}
  \bibinfo{author}{\bibfnamefont{D.-S.} \bibnamefont{Sun}},
  \bibinfo{journal}{Physica Scripta} \textbf{\bibinfo{volume}{76}},
  \bibinfo{pages}{428} (\bibinfo{year}{2007}).

\bibitem[{\citenamefont{Feng et~al.}(2010{\natexlab{b}})\citenamefont{Feng,
  Kaplinghat, and Yu}}]{Feng:2009hw}
\bibinfo{author}{\bibfnamefont{J.~L.} \bibnamefont{Feng}},
  \bibinfo{author}{\bibfnamefont{M.}~\bibnamefont{Kaplinghat}},
  \bibnamefont{and} \bibinfo{author}{\bibfnamefont{H.-B.} \bibnamefont{Yu}},
  \bibinfo{journal}{Phys. Rev. Lett.} \textbf{\bibinfo{volume}{104}},
  \bibinfo{pages}{151301} (\bibinfo{year}{2010}{\natexlab{b}}),
  \eprint{0911.0422}.

\bibitem[{\citenamefont{Yuan et~al.}(2009)}]{Yuan:2009bb}
\bibinfo{author}{\bibfnamefont{Q.}~\bibnamefont{Yuan}} \bibnamefont{et~al.},
  \bibinfo{journal}{JCAP} \textbf{\bibinfo{volume}{0912}}, \bibinfo{pages}{011}
  (\bibinfo{year}{2009}), \eprint{0905.2736}.

\bibitem[{\citenamefont{Dent et~al.}(2010)\citenamefont{Dent, Dutta, and
  Scherrer}}]{Dent:2009bv}
\bibinfo{author}{\bibfnamefont{J.~B.} \bibnamefont{Dent}},
  \bibinfo{author}{\bibfnamefont{S.}~\bibnamefont{Dutta}}, \bibnamefont{and}
  \bibinfo{author}{\bibfnamefont{R.~J.} \bibnamefont{Scherrer}},
  \bibinfo{journal}{Phys. Lett.} \textbf{\bibinfo{volume}{B687}},
  \bibinfo{pages}{275} (\bibinfo{year}{2010}), \eprint{0909.4128}.

\bibitem[{\citenamefont{Zavala et~al.}(2010)\citenamefont{Zavala, Vogelsberger,
  and White}}]{Zavala:2009mi}
\bibinfo{author}{\bibfnamefont{J.}~\bibnamefont{Zavala}},
  \bibinfo{author}{\bibfnamefont{M.}~\bibnamefont{Vogelsberger}},
  \bibnamefont{and} \bibinfo{author}{\bibfnamefont{S.~D.~M.}
  \bibnamefont{White}}, \bibinfo{journal}{Phys. Rev.}
  \textbf{\bibinfo{volume}{D81}}, \bibinfo{pages}{083502}
  (\bibinfo{year}{2010}), \eprint{0910.5221}.

\bibitem[{\citenamefont{Xue et~al.}(2008)}]{Xue:2008se}
\bibinfo{author}{\bibfnamefont{X.~X.} \bibnamefont{Xue}} \bibnamefont{et~al.}
  (\bibinfo{collaboration}{SDSS}), \bibinfo{journal}{Astrophys. J.}
  \textbf{\bibinfo{volume}{684}}, \bibinfo{pages}{1143} (\bibinfo{year}{2008}),
  \eprint{0801.1232}.

\bibitem[{\citenamefont{Reid et~al.}(2009)\citenamefont{Reid, Menten,
  Brunthaler, and Moellenbrock}}]{Reid:2009nr}
\bibinfo{author}{\bibfnamefont{M.~J.} \bibnamefont{Reid}},
  \bibinfo{author}{\bibfnamefont{K.~M.} \bibnamefont{Menten}},
  \bibinfo{author}{\bibfnamefont{A.}~\bibnamefont{Brunthaler}},
  \bibnamefont{and} \bibinfo{author}{\bibfnamefont{G.~A.}
  \bibnamefont{Moellenbrock}} (\bibinfo{year}{2009}), \eprint{0902.3928}.

\bibitem[{\citenamefont{Smith et~al.}(2007)}]{Smith:2006ym}
\bibinfo{author}{\bibfnamefont{M.~C.} \bibnamefont{Smith}}
  \bibnamefont{et~al.}, \bibinfo{journal}{Mon. Not. Roy. Astron. Soc.}
  \textbf{\bibinfo{volume}{379}}, \bibinfo{pages}{755} (\bibinfo{year}{2007}),
  \eprint{astro-ph/0611671}.

\bibitem[{\citenamefont{Kuhlen and Malyshev}(2009)}]{Kuhlen:2009is}
\bibinfo{author}{\bibfnamefont{M.}~\bibnamefont{Kuhlen}} \bibnamefont{and}
  \bibinfo{author}{\bibfnamefont{D.}~\bibnamefont{Malyshev}},
  \bibinfo{journal}{Phys. Rev.} \textbf{\bibinfo{volume}{D79}},
  \bibinfo{pages}{123517} (\bibinfo{year}{2009}), \eprint{0904.3378}.

\bibitem[{\citenamefont{Cline et~al.}(2010)\citenamefont{Cline, Vincent, and
  Xue}}]{Cline:2010ag}
\bibinfo{author}{\bibfnamefont{J.~M.} \bibnamefont{Cline}},
  \bibinfo{author}{\bibfnamefont{A.~C.} \bibnamefont{Vincent}},
  \bibnamefont{and} \bibinfo{author}{\bibfnamefont{W.}~\bibnamefont{Xue}},
  \bibinfo{journal}{Phys. Rev.} \textbf{\bibinfo{volume}{D81}},
  \bibinfo{pages}{083512} (\bibinfo{year}{2010}), \eprint{1001.5399}.

\bibitem[{\citenamefont{Stawarz et~al.}(2010)\citenamefont{Stawarz, Petrosian,
  and Blandford}}]{Stawarz:2009ig}
\bibinfo{author}{\bibfnamefont{L.}~\bibnamefont{Stawarz}},
  \bibinfo{author}{\bibfnamefont{V.}~\bibnamefont{Petrosian}},
  \bibnamefont{and} \bibinfo{author}{\bibfnamefont{R.~D.}
  \bibnamefont{Blandford}}, \bibinfo{journal}{Astrophys. J.}
  \textbf{\bibinfo{volume}{710}}, \bibinfo{pages}{236} (\bibinfo{year}{2010}),
  \eprint{0908.1094}.

\bibitem[{\citenamefont{Katz et~al.}(2009)\citenamefont{Katz, Blum, and
  Waxman}}]{Katz:2009yd}
\bibinfo{author}{\bibfnamefont{B.}~\bibnamefont{Katz}},
  \bibinfo{author}{\bibfnamefont{K.}~\bibnamefont{Blum}}, \bibnamefont{and}
  \bibinfo{author}{\bibfnamefont{E.}~\bibnamefont{Waxman}}
  (\bibinfo{year}{2009}), \eprint{0907.1686}.

\bibitem[{\citenamefont{Bringmann et~al.}(2009)\citenamefont{Bringmann, Doro,
  and Fornasa}}]{Bringmann:2008kj}
\bibinfo{author}{\bibfnamefont{T.}~\bibnamefont{Bringmann}},
  \bibinfo{author}{\bibfnamefont{M.}~\bibnamefont{Doro}}, \bibnamefont{and}
  \bibinfo{author}{\bibfnamefont{M.}~\bibnamefont{Fornasa}},
  \bibinfo{journal}{JCAP} \textbf{\bibinfo{volume}{0901}}, \bibinfo{pages}{016}
  (\bibinfo{year}{2009}), \eprint{0809.2269}.

\bibitem[{\citenamefont{Padmanabhan and Finkbeiner}(2005)}]{Padmanabhan:2005es}
\bibinfo{author}{\bibfnamefont{N.}~\bibnamefont{Padmanabhan}} \bibnamefont{and}
  \bibinfo{author}{\bibfnamefont{D.~P.} \bibnamefont{Finkbeiner}},
  \bibinfo{journal}{Phys. Rev.} \textbf{\bibinfo{volume}{D72}},
  \bibinfo{pages}{023508} (\bibinfo{year}{2005}), \eprint{astro-ph/0503486}.

\bibitem[{\citenamefont{Galli et~al.}(2009)\citenamefont{Galli, Iocco, Bertone,
  and Melchiorri}}]{Galli:2009zc}
\bibinfo{author}{\bibfnamefont{S.}~\bibnamefont{Galli}},
  \bibinfo{author}{\bibfnamefont{F.}~\bibnamefont{Iocco}},
  \bibinfo{author}{\bibfnamefont{G.}~\bibnamefont{Bertone}}, \bibnamefont{and}
  \bibinfo{author}{\bibfnamefont{A.}~\bibnamefont{Melchiorri}},
  \bibinfo{journal}{Phys. Rev.} \textbf{\bibinfo{volume}{D80}},
  \bibinfo{pages}{023505} (\bibinfo{year}{2009}), \eprint{0905.0003}.

\bibitem[{\citenamefont{Slatyer et~al.}(2009)\citenamefont{Slatyer,
  Padmanabhan, and Finkbeiner}}]{Slatyer:2009yq}
\bibinfo{author}{\bibfnamefont{T.~R.} \bibnamefont{Slatyer}},
  \bibinfo{author}{\bibfnamefont{N.}~\bibnamefont{Padmanabhan}},
  \bibnamefont{and} \bibinfo{author}{\bibfnamefont{D.~P.}
  \bibnamefont{Finkbeiner}} (\bibinfo{year}{2009}), \eprint{0906.1197}.

\bibitem[{\citenamefont{Galli et~al.}(2010)}]{Galli:2010it}
\bibinfo{author}{\bibfnamefont{S.}~\bibnamefont{Galli}} \bibnamefont{et~al.}
  (\bibinfo{year}{2010}), \eprint{1005.3808}.

\bibitem[{\citenamefont{Albert et~al.}(2008)}]{Albert2008}
\bibinfo{author}{\bibfnamefont{J.}~\bibnamefont{Albert}} \bibnamefont{et~al.}
  (\bibinfo{collaboration}{MAGIC}), \bibinfo{journal}{Astrophys. J.}
  \textbf{\bibinfo{volume}{679}}, \bibinfo{pages}{428} (\bibinfo{year}{2008}),
  \eprint{0711.2574}.

\bibitem[{\citenamefont{Acciari et~al.}(2010)}]{Veritas:2010pj}
\bibinfo{author}{\bibfnamefont{V.}~\bibnamefont{Acciari}} \bibnamefont{et~al.}
  (\bibinfo{collaboration}{The VERITAS}) (\bibinfo{year}{2010}),
  \eprint{1006.5955}.

\bibitem[{\citenamefont{Willman}(2009)}]{Willman:2009dv}
\bibinfo{author}{\bibfnamefont{B.}~\bibnamefont{Willman}}
  (\bibinfo{year}{2009}), \eprint{0907.4758}.

\end{thebibliography}

\end{document}